\documentclass{article}
\usepackage{float}
\usepackage[preprint]{neurips_2024}
\usepackage[utf8]{inputenc}
\usepackage[T1]{fontenc}
\usepackage{graphicx}
\usepackage{amsmath}
\usepackage{amssymb}
\usepackage{svg}
\usepackage{booktabs}
\usepackage{multirow}
\PassOptionsToPackage{table}{xcolor}
\usepackage[most]{tcolorbox}
\usepackage{colortbl}
\usepackage{makecell}
\usepackage{hyperref}
\usepackage{fontawesome5}
\usepackage{tabularx}

\definecolor{rowpink}{RGB}{253, 242, 242}

\title{
  \Large
  \raisebox{-0.8ex}{
    \includegraphics[height=2.8ex]{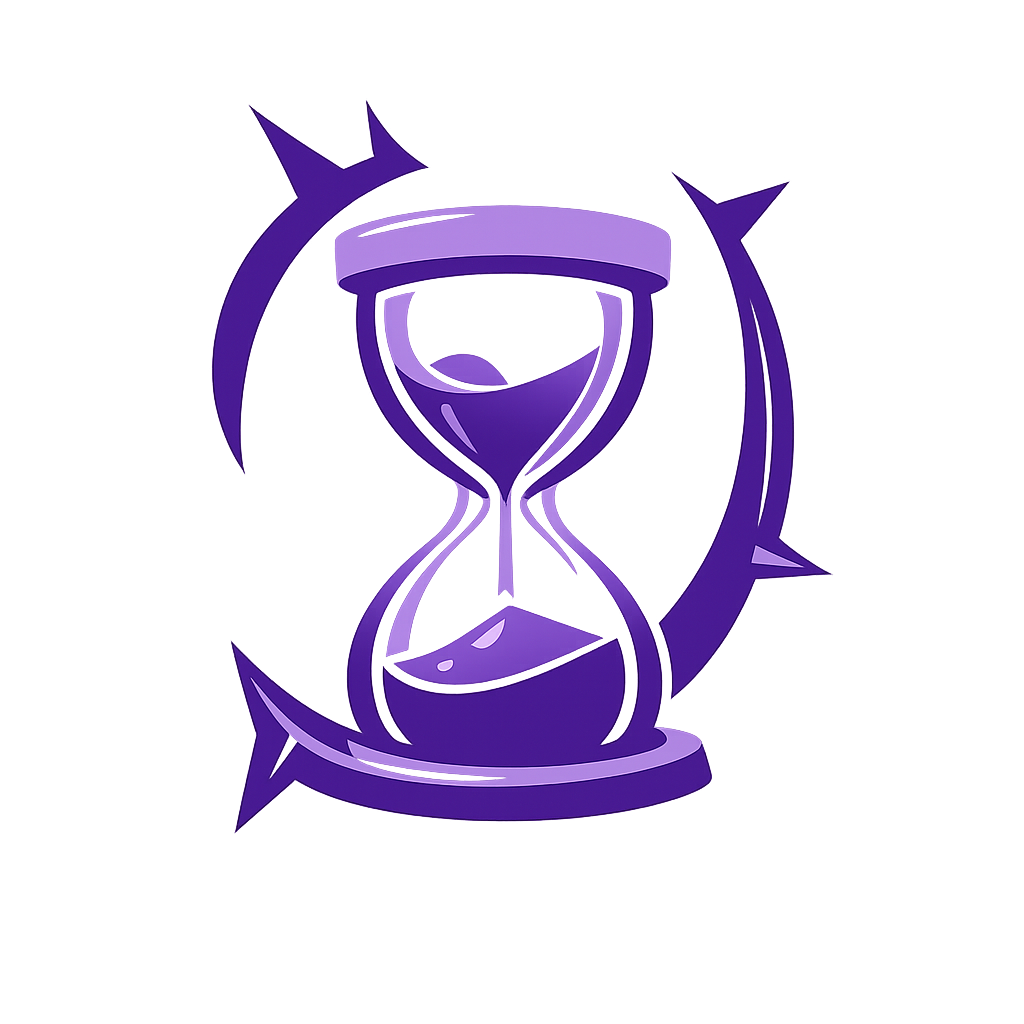}
  }
  \hspace{-0.5em}
  Eureka-Audio: Triggering Audio Intelligence in Compact Language Models
}

\author{\vspace{-4ex}}

\begin{document}

\maketitle

\begin{center}
  \vspace{-6.0em}

  \textbf{Dan Zhang}$^{*1}$ \hspace{1.2em}
  \textbf{Yishu Lei}$^{*1}$ \hspace{1.2em}
  \textbf{Jing Hu}$^{*1, 3}$ \hspace{1.2em}
  \textbf{Shuwei He}$^{*1, 2}$ \hspace{1.2em}
  \textbf{Songhe Deng}$^{1}$
  \\[0.3em]

  \textbf{Xianlong Luo}$^{1}$ \hspace{1.2em}
  \textbf{Danxiang Zhu}$^{1}$ \hspace{1.2em}
  \textbf{Shikun Feng}$^{\dagger1}$ \hspace{1.2em}
  \textbf{Rui Liu}$^{2}$
  \\[0.3em]

  \textbf{Jingzhou He}$^{1}$ \hspace{1.2em}
  \textbf{Yu Sun}$^{1}$ \hspace{1.2em}
  \textbf{Hua Wu}$^{1}$ \hspace{1.2em}
  \textbf{Haifeng Wang}$^{1}$
  \\[0.8em]

  { \normalfont \small
    $^{1}$Baidu Inc. \qquad
    $^{2}$College of Computer Science, Inner Mongolia University \\
    $^{3}$Tsinghua Shenzhen International Graduate School, Tsinghua University
  }
  \\[0.6em]

  { \normalfont \footnotesize \ttfamily
    \{zhangdan20, leiyishu, heshuwei, dengsonghe, luoxianlong, zhudanxiang\}@baidu.com \\
    \{fengshikun01, hejingzhou, sunyu02, wu\_hua, wanghaifeng\}@baidu.com \\
    cminusser@gmail.com, \quad imucslr@imu.edu.cn
  }
\end{center}

{
  \renewcommand{\thefootnote}{}
  \footnotetext{\noindent \kern-1.1em 
    \begin{tabular}{@{}l}
      $^*$Equal contribution. \hspace{1.5em} $^\dagger$Corresponding author. \\
      Project Page: \href{https://github.com/Alittleegg/Eureka-Audio}{\texttt{https://github.com/Alittleegg/Eureka-Audio}}
    \end{tabular}
  }
}

\vspace{-1em}

\begin{figure*}[ht!]
  \centering
  \includegraphics[width=\textwidth]{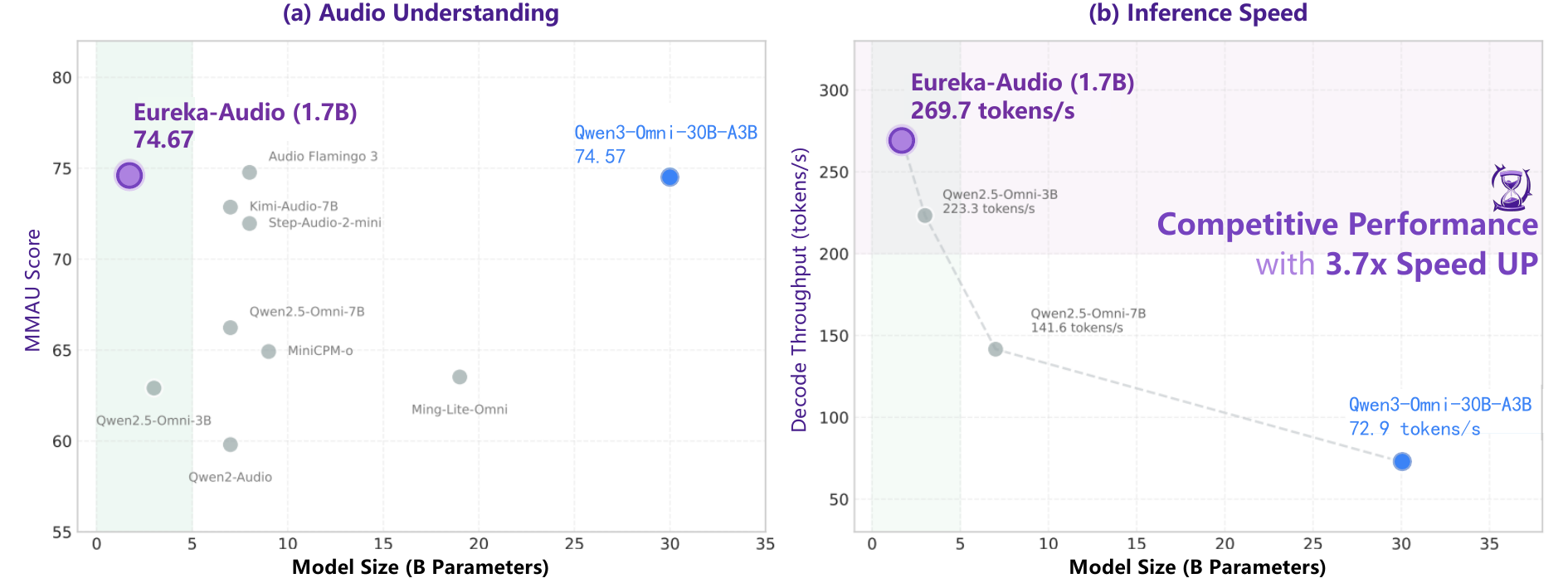}
  \caption{Comparison of Eureka-Audio with open-source audio-language and omni-modal baselines. (a) On the MMAU
  benchmark, Eureka-Audio (1.7B) achieves a score of 74.67, competitive with models 4--17$\times$ larger. (b) Eureka-Audio
  achieves the highest decode throughput of 269.7 tokens/sec among the compared models.}
  \label{fig:teaser}
\end{figure*}

\begin{abstract}
We present \textbf{Eureka-Audio}, a compact yet high-performance audio language model that achieves competitive performance against models that are 4 to 18 times larger across a broad range of audio understanding benchmarks. Despite containing only 1.7B parameters, Eureka-Audio demonstrates strong performance on automatic speech recognition (ASR), audio understanding, and dense audio captioning, matching or surpassing multiple 7B to 30B audio and omni-modal baselines. The model adopts a unified end-to-end architecture composed of a lightweight language backbone, a Whisper-based audio encoder, and a sparsely activated Mixture-of-Experts (MoE) adapter that explicitly accounts for audio heterogeneity and alleviates cross-modal optimization conflicts under limited capacity. To further enhance paralinguistic reasoning, we introduce DataFlux, a closed loop audio instruction data synthesis and verification pipeline that constructs high quality, logically consistent supervision from raw audio. Extensive evaluations across ASR, knowledge reasoning, safety, instruction following, and paralinguistic benchmarks, demonstrate that Eureka-Audio achieves an efficient balance between computational cost and performance. These results establish Eureka Audio as a strong and practical baseline for lightweight audio understanding models.
\end{abstract}

\section{Introduction}
Recent advances in multimodal large language models have driven a paradigm shift in the audio domain, moving beyond speech recognition toward more general audio understanding. Unlike conventional speech-centric approaches that focus on transcription, audio understanding requires models to jointly capture semantic content (e.g., what is said or what event occurs) and paralinguistic cues (e.g., emotion, tone, laughter, hesitation, and emphasis). Such capabilities are crucial for real-world applications including intelligent assistants, customer service quality inspection, content retrieval, and general-purpose audio analysis.

Despite this progress, deploying high-quality audio understanding models in practical settings remains challenging. Recent methods often rely on substantially larger models, such as Kimi-Audio-7B~\cite{Kimi-Audio}, Step-Audio-2-mini-8B~\cite{Step-audio-2}, Qwen2-Audio-7B~\cite{Qwen2-audio}, and Qwen3-Omni-30B-A3B-Instruct~\cite{qwen3-omni}, to achieve strong performance. 
However, the high inference latency and computational costs overhead inherent in these models often preclude their use in real-time or resource-constrained scenarios.

Consequently, there is still a notable scarcity of lightweight, open-source audio understanding models that offer both high performance and full reproducibility.

Achieving strong audio understanding under a lightweight setting is non-trivial. Audio signals are inherently heterogeneous, with speech, environmental sounds, and music exhibiting distinct statistical structures and representational characteristics. Naively using a single shared projection for cross-modal alignment often introduces conflicting optimization signals and limits representational capacity. Moreover, under limited model capacity, semantic information and paralinguistic cues are more prone to competing for limited representational capacity, leading to reduced parameter efficiency and degraded generalization. These challenges are further exacerbated by the scarcity of high-quality audio instruction and preference data, which can cause instability and performance degradation during post-training.

To address these challenges, we introduce \textbf{Eureka-Audio}, an open-source, high-performance lightweight model for audio semantic and paralinguistic understanding. Eureka-Audio adopts Qwen3-1.7B-base~\cite{Qwen3} as its language backbone. Audio inputs are first encoded into continuous acoustic representations using a Whisper-based audio encoder, and are then mapped into the language model’s semantic space through a sparsely activated Mixture-of-Experts (MoE) adapter~\cite{moe-adapter}. This sparse adaptation mechanism explicitly accounts for audio heterogeneity and improves cross-modal alignment quality.

We further construct a closed loop training pipeline covering both pre-training and post-training stages. During pre-training, the model acquires fundamental audio understanding and cross-modal alignment capabilities. During post-training, we combine open-source and in-house data and introduce \textbf{DataFlux}, a systematic audio instruction data synthesis pipeline that enables structured data management and continuous updates.

As shown in Figure~\ref{fig:teaser}, Eureka-Audio consistently achieves competitive performance across ASR, audio semantic understanding, and paralinguistic understanding benchmarks, despite being significantly smaller than many strong audio and omni-modal baselines, while also delivering up to 3.7$\times$ faster decoding speed.

Our contributions are summarized as follows:

\begin{itemize}
    \item \textbf{Eureka-Audio is introduced as a lightweight yet high-performance audio understanding model} with only \textbf{1.7B parameters}, achieving strong results on both audio semantic and paralinguistic understanding tasks while remaining suitable for efficient real-world deployment.
    
    \item \textbf{A sparsely activated MoE-based adapter} is developed between the Whisper-based audio encoder and the language backbone to explicitly address audio heterogeneity and improve cross-modal alignment. Based on this design, a complete training pipeline 

    significantly enhances training stability and multimodal alignment quality.
    
    \item \textbf{DataFlux is a structured audio instruction data synthesis pipeline} designed to systematically construct high-quality paralinguistic instruction data. Through a multi-step data generation and validation process, the pipeline ensures reliable semantic alignment and logical consistency of the synthesized data, thereby supporting the post-training stage and enabling models to progressively enhance audio paralinguistic understanding and reasoning under controlled, high-quality data supervision.
    
    \item \textbf{A novel evaluation methodology for audio captioning is proposed}, enabling more faithful assessment of high-level audio understanding across diverse audio types, including speech, sound, and music.
\end{itemize}

\section{Related Work}
\subsection{Large Audio Language Model}
Existing Large Audio Language Models~\citep{chu2023qwenaudioadvancinguniversalaudio, Qwen2-audio, Step-audio-2, Kimi-Audio, zeng2024glm, fang2024llamaomniseamlessspeechinteraction,ai2025ming} typically adopt an end-to-end architecture built upon a pre-trained LLM backbone. 
In this framework, acoustic features are integrated into the model via an audio encoder, with a trainable adapter acting as the bridge between the encoder and the LLM. 
Most existing works, including ours, utilize Automatic Speech Recognition (ASR) models, such as Whisper~\citep{radford2023robust}, as the audio encoder due to their robust representation capabilities. In contrast, a subset of methods~\citep{Step-audio-2, Kimi-Audio} involves using an audio tokenizer, such as Residual Vector Quantization (RVQ), or Finite Scalar Quantization(FSQ), to quantize audio signals into discrete codebook~\citep{mentzer2023finite} indices for model input.

\subsection{Lightweight Multimodal Models}

Since Large Audio Language Models prioritize real-time interactivity, they demand strictly optimized FLOPs and inference latency. While mainstream models~\citep{chu2023qwenaudioadvancinguniversalaudio, Qwen2-audio, Step-audio-2, Kimi-Audio, zeng2024glm, fang2024llamaomniseamlessspeechinteraction} typically exceed 7B parameters, leading to substantial computational overhead, some recent works~\citep{ai2025ming,qwen3-omni} adopt MoE architectures to reduce active parameters to approximately 3B. However, the excessive total memory footprint of MoE models remains a significant bottleneck for deployment on resource-constrained edge devices. To address these challenges, we propose a 1.7B dense model, which significantly alleviates the deployment overhead while maintaining high-performance audio interaction.

\section{Architecture}
\begin{figure*}[htb!]
  \centering
  \includegraphics[width=1.0\textwidth]{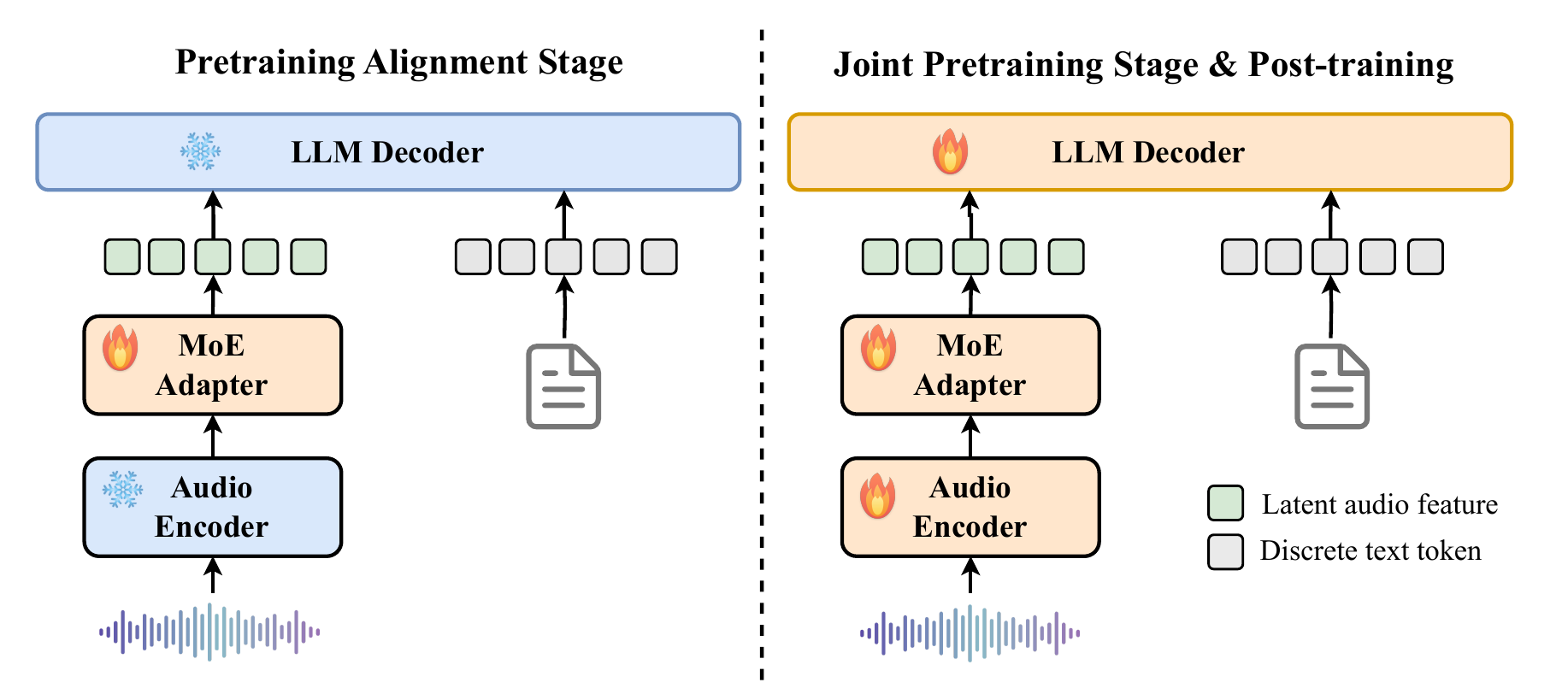}
  \caption{The overview of the Eureka-Audio. Eureka-Audio adopts a unified end-to-end design consisting of three core components: (1) \textbf{a Whisper-based audio encoder} that encodes raw waveforms into high–temporal-resolution acoustic representations; (2) \textbf{a sparse MoE adapter}~\cite{moe-adapter} that maps acoustic features into the language model embedding space for efficient cross-modal alignment; and (3) \textbf{a lightweight language model backbone} (Qwen3-1.7B-base~\cite{Qwen3}) that jointly models aligned audio embeddings and text tokens in an autoregressive manner to support diverse audio understanding tasks.}
  \label{fig:architecture}
\end{figure*}

\subsection{Overview}
We propose Eureka-Audio, a lightweight audio–language model designed for general audio understanding tasks. The model adopts a unified end-to-end architecture that tightly integrates Whisper-based audio encoder, a sparse MoE adapter, and a lightweight language model backbone, achieving a favorable balance between modeling capacity and computational and parameter efficiency. As illustrated in Figure~\ref{fig:architecture}, Eureka-Audio consists of three core components:

\paragraph{Audio Encoder}
we employ a Whisper~\cite{whisper}-based audio encoder to encode raw audio waveforms into high–temporal-resolution acoustic features, which capture fine-grained perceptual and semantic information present in the audio signal.

\paragraph{Sparse MoE Adapter.}
The continuous acoustic features produced by the Whisper~\cite{whisper}-based audio encoder are first fed into a MoE Adapter~\cite{moe-adapter}, which maps audio representations into the embedding space of the language model. Serving as a critical interface between audio representations and the language backbone, the MoE Adapter enables efficient cross-modal alignment through sparse expert routing. This design explicitly models the heterogeneity of audio signals at both the semantic and acoustic levels, mitigating optimization conflicts while improving representational efficiency and modeling flexibility under controlled parameter and computational overhead.

\paragraph{Language Model Backbone.}
We adopt Qwen3-1.7B-base~\cite{Qwen3} as the language model backbone. After alignment via the MoE Adapter, audio embeddings are concatenated with text token embeddings along the sequence dimension and jointly modeled by the backbone in a standard autoregressive manner. The model outputs text tokens, enabling a wide range of downstream audio understanding tasks, including audio question answering and instruction following.

Overall, this architecture allows Eureka-Audio to perform both audio semantic understanding and paralinguistic reasoning within a unified and lightweight framework, while remaining well suited for deployment in resource-constrained scenarios.

\subsection{Sparse MoE Adapter}

To project audio representations into the LLM embedding space, we adopt a sparse Mixture-of-Experts (MoE) adapter~\cite{moe-adapter} instead of a conventional dense projector.
Given an input audio token $\mathbf{x} \in \mathbb{R}^d$, a learnable router computes gating logits $G(\mathbf{x})=\mathbf{x}\mathbf{W}_g$ and selects the Top-$k$ experts via sparse softmax routing.
Each expert is implemented as a lightweight feed-forward network with SiLU activation.
The selected expert outputs are aggregated according to the routing weights and further mapped to the LLM embedding dimension through a linear projection followed by layer normalization:
\begin{equation}
\mathbf{y}_{\text{MoE}} =
\mathcal{LN}\!\left(
\mathbf{W}_P
\sum_{i \in \mathcal{I}} G(\mathbf{x})_i \cdot E_i(\mathbf{x})
\right),
\end{equation}
where $\mathcal{I}$ denotes the set of routed experts and $G(\mathbf{x})$ is the sparse gating distribution.

The resulting adapted audio embeddings $\mathbf{Y}_{\text{MoE}} = \{\mathbf{y}_{\text{MoE}}^{(1)}, \dots, \mathbf{y}_{\text{MoE}}^{(T_a)}\}$ are concatenated with text token embeddings to form the final input sequence to the LLM.

\paragraph{Training Objective.}

Let $\mathbf{Z}$ denote the complete input embedding sequence to the LLM, consisting of the adapted audio embeddings and textual token embeddings.
The model is trained end-to-end using the standard autoregressive next-token prediction (NTP) objective:
\begin{equation}
\mathcal{L}_{\text{NTP}}
=
- \sum_{t=1}^{T}
\log P(y_t \mid y_{<t}, \mathbf{Z}; \theta),
\end{equation}
where $\theta$ denotes all trainable parameters including the MoE adapter and the LLM backbone.

To mitigate expert collapse, we incorporate a load-balancing auxiliary loss over the routed experts.
Let $B$ be the number of routed audio tokens in a batch, and $p_{b,e}$ denote the routing probability of token $b$ to expert $e$.
The auxiliary loss is defined as:
\begin{equation}
\mathcal{L}_{\text{aux}}
=
|\mathcal{E}_R|
\sum_{e \in \mathcal{E}_R}
\bar{P}_e \cdot \bar{f}_e,
\end{equation}
where
$\bar{P}_e=\frac{1}{B}\sum_{b=1}^{B}p_{b,e}$ represents the expert importance and
$\bar{f}_e$ is the fraction of tokens routed to expert $e$.
The final objective is:
\begin{equation}
\mathcal{L}
=
\mathcal{L}_{\text{NTP}}
+
\lambda \mathcal{L}_{\text{aux}}.
\end{equation}

\section{Pretraining}
We adopt a two-stage pretraining framework for Eureka-Audio, consisting of an alignment stage~(Stage 1) and a joint pretraining stage~(Stage 2). 
This design aims to first establish a stable audio--text modality alignment and subsequently achieve comprehensive audio understanding through the joint optimization of the full model.

\paragraph{Stage 1~(Alignment Stage).}
In this stage, only the \textbf{MoE Adapter} is trainable, while the parameters of both the language model backbone and the audio encoder remain frozen. 
Training leverages three types of data: audio unimodal modeling, audio-to-text mapping, and audio--text interleaved modeling. 
This configuration enables the adapter to learn an effective mapping from acoustic representations to the embedding space of the language model, establishing a robust audio--text modality alignment. 
The primary objective of this stage is to ensure high-quality cross-modal alignment and training stability.

\paragraph{Stage 2~(Joint Pretraining Stage).}
In this stage, all model parameters are unfrozen and jointly optimized. In addition to the data types utilized in Stage~1, we incorporate audio captioning data to enhance the capacity of the model to capture high-level semantic information and paralinguistic cues. This stage enables the model to learn richer audio semantics and complex reasoning patterns, leading to improved overall audio understanding capabilities.

\subsection{Task Formulation}
\label{sec:task_formulation}

Our pretraining objective is formulated over a mixture of tasks, including:

\begin{itemize}
    \item \textbf{Unimodal Modeling.}
    This category involves next-token prediction for both text tokens and discrete semantic audio tokens\footnote{The discrete semantic audio tokens are extracted using the tokenizer from GLM-4-Voice~\cite{glm4voice}.}. These tasks serve to preserve the language modeling capability of the backbone and to learn the fundamental distribution of audio representations.

    \item \textbf{Audio--Text Mapping.}
    This category encompasses automatic speech recognition~(ASR, Audio $\rightarrow$ Text) and text-to-speech~(TTS, Text $\rightarrow$ Audio semantic tokens), providing direct supervision for the cross-modal correspondence between audio and text.

    \item \textbf{Audio--Text Interleaving.}
    We construct interleaved sequences under audio-conditioned settings, where the model is trained to predict either the next semantic audio token or the next text token. This task further strengthens the coupling between acoustic and linguistic representations.

    \item \textbf{Audio Captioning.}
    This category covers tasks such as acoustic scene understanding, emotion recognition, sound event detection, sound source identification, and music understanding, offering high-level semantic supervision for holistic audio understanding.
\end{itemize}

\subsection{Dataset Composition and Distribution}
\label{sec:dataset_composition_and_distribution}

During Stage~1 (Alignment Stage), the model is trained on approximately \textbf{100B tokens}. In Stage~2 (Joint Pretraining Stage), the total training scale scales up to approximately \textbf{1T tokens}. The approximate scale of each data category is summarized in Table~\ref{tab:data_distribution}. Notably, audio captioning data is introduced exclusively in Stage~2 and consists of various open-source datasets; detailed specifications are provided in Appendix~\ref{app:opensource_audio_for_pretraining}.

\begin{table}[h!]
\centering
\small
\caption{Data distribution and training schedule across different training tasks and stages. Audio and text data are jointly sampled with a fixed 1:1 ratio. For brevity, the table reports statistics for the \textbf{audio modality only}.}
\label{tab:data_distribution}
\resizebox{\linewidth}{!}{%
    \begin{tabular}{lcccc}
    \toprule
    \textbf{Task} & \textbf{Audio Length (Hours)} & \textbf{Tokens (B)} & \textbf{Stage~1 Task Ratio} & \textbf{Stage~2 Task Ratio} \\
    \midrule
    Audio Unimodal Modeling   & 500{,}000 & 25  & 0.2  & 0.03 \\
    \addlinespace
    Audio--Text Mapping       & 5{,}500{,}000 & 360 & 0.45 & 0.56 \\
    \addlinespace
    Audio--Text Interleaving  & 5{,}150{,}000 & 150 & 0.35 & 0.07 \\
    \addlinespace
    Audio Captioning          & 220{,}000 & 18  & --   & 0.34 \\
    \bottomrule
    \end{tabular}%
} 
\end{table}

\section{Post-training}
\subsection{DataFlux}
\label{sec:dataflux}

\begin{figure*}[h!]
  \centering
  \includegraphics[width=\linewidth]{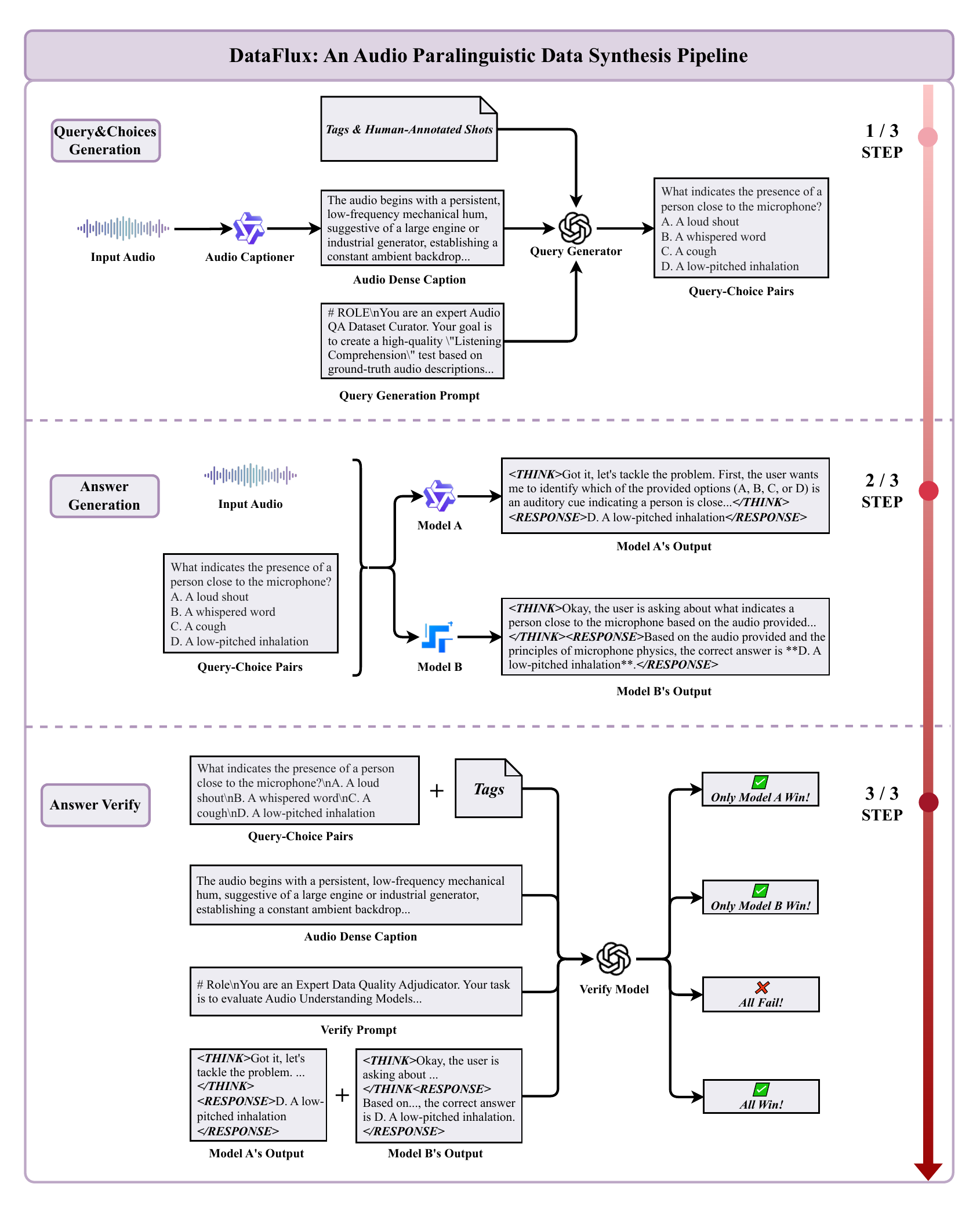}
  \caption{Overview of DataFlux. Starting from raw audio, DataFlux constructs high-quality paralinguistic instruction data through a three-step workflow: 
    (1) \textbf{Query--Choice Generation}, where dense audio captions are first produced and then transformed into structured Query--Choice pairs using a predefined paralinguistic taxonomy and few-shot exemplars; 
    (2) \textbf{Answer Generation}, where multiple audio large language models generate reasoning traces and answers conditioned on the same audio and queries; and 
    (3) \textbf{Answer Verification}, where an automated judge evaluates multi-model outputs based on logical consistency and alignment with the audio content, retaining reliable samples while filtering noisy or inconsistent ones.}
  \label{fig:dataflux}
\end{figure*}

To systematically construct high-quality audio paralinguistic instruction data and effectively facilitate paralinguistic understanding and reasoning during post-training, we propose \textbf{DataFlux}, a data synthesis and filtering pipeline tailored for paralinguistic tasks. Existing open-source post-training datasets often derive from earlier model versions, potentially leading to misalignment with the requirements of more capable models as their reasoning abilities evolve. Starting from raw audio, DataFlux progressively builds structured, logically consistent, and reasoning-oriented audio instruction data through a multi-stage process of generation, alignment, and validation.

As illustrated in Figure~\ref{fig:dataflux}, DataFlux operates via a three-step workflow. In \textbf{Step~1}, raw audio is processed by an audio captioner (Qwen3-Omni-30B-A3B-Captioner~\cite{qwen3-omni}) to generate Audio Dense Captions, which provide fine-grained descriptions of acoustic events, environmental contexts, and potential paralinguistic cues. Based on a predefined paralinguistic taxonomy and a small set of manually curated exemplars, a large language model (GPT-OSS-120B~\cite{gptoss}) transforms these captions into structured instruction formats, producing Query--Choice pairs strictly aligned with the audio content. This step effectively maps continuous, unstructured audio signals into a discrete instruction space suitable for model training.

In \textbf{Step~2}, the generated Query--Choice pairs, along with the original audio, are fed into multiple audio large language models (Qwen3-Omni-30B-A3B-Thinking~\cite{qwen3-omni} and Step-Audio-R1~\cite{Step-Audio-R1}) to produce reasoning traces and final answers. By leveraging models with distinct reasoning characteristics, this step explicitly induces answer diversity, serving as a foundation for subsequent data filtering and hard-example mining. The outputs include not only final answers but also intermediate reasoning traces, enabling a more granular assessment of paralinguistic understanding.

In \textbf{Step~3}, DataFlux employs an automated judging mechanism to systematically evaluate the multi-model outputs. A judge model (GPT-OSS-120B~\cite{gptoss}) jointly assesses the original audio captions, the Query--Choice pairs, and the reasoning traces and answers produced by the different models. Judgments rely on logical consistency, coverage of salient details, and the absence of semantic conflicts with audio descriptions. Based on these evaluations, samples are categorized by quality: those exhibiting consistent reasoning and high agreement with the audio captions are retained, while instances with clear conflicts or reasoning failures are filtered out, thereby effectively reducing noise during post-training.

Through this three-step design, DataFlux realizes an end-to-end automated pipeline for constructing high-quality paralinguistic instruction data from raw audio. The pipeline ensures reliable semantic alignment and logical consistency of the synthesized data, providing a scalable and extensible source for controlled, high-quality post-training supervision.

\subsection{Supervised Fine-Tuning}
\label{sec:sft}
During the Supervised Fine-Tuning~(SFT) stage, all model parameters are unfrozen and jointly optimized, with a total training scale of approximately \textbf{30B tokens}. Throughout training, text and audio modalities are sampled using a fixed 1:1 ratio. The detailed scale and composition of the audio data are summarized in Table~\ref{tab:sft_data_composition}.

\begin{table}[h!]
\centering
\caption{Audio data composition and sampling ratios used during the SFT stage.}
\label{tab:sft_data_composition}
\begin{tabular}{lcc}
\toprule
\textbf{Task Type} & \textbf{Audio Length (Hours)} & \textbf{Ratio} \\
\midrule
ASR                          & 250{,}000 & 0.6 \\
Paralinguistic Understanding & 2{,}500   & 0.1 \\
Semantic Understanding       & 100{,}000 & 0.2 \\
Audio Dense Captioning       & 2{,}500   & 0.1 \\
\bottomrule
\end{tabular}
\end{table}

\section{Evaluation}
In this section, we conduct a comprehensive evaluation of Eureka-Audio, covering automatic speech recognition (ASR), audio understanding, and dense audio captioning, to assess its effectiveness as a lightweight audio understanding model. Unlike conventional evaluation protocols that primarily focus on transcription accuracy or isolated audio classification tasks, our evaluation emphasizes high-level audio understanding, requiring models to jointly reason over semantic content and paralinguistic cues across diverse audio types, including speech, sounds, and music.

We compare Eureka-Audio with a diverse set of large scale audio language models and omni modal language models, and report results under their officially recommended inference configurations to ensure fair comparison. Despite having only 1.7 billion parameters of LLM backbone, Eureka-Audio consistently achieves competitive performance across a wide range of benchmarks, demonstrating an efficient balance between model capacity and task performance. Furthermore, we conduct systematic ablation studies to examine the contribution of the proposed DataFlux framework and to quantify its impact on overall audio understanding performance.

\subsection{Automatic Speech Recognition}
To evaluate the automatic speech recognition (ASR) capability of Eureka-Audio, we conduct a systematic evaluation on a diverse set of standard benchmarks covering multiple languages and acoustic conditions, including LibriSpeech~\cite{librispeech}, Fleurs~\cite{fleurs}, AISHELL~\cite{aishell-2}, and WenetSpeech~\cite{wenetspeech}. These benchmarks span both English and Mandarin speech, and cover a wide range of scenarios such as clean speech, noisy environments, conversational speech, and meeting-style recordings. We report word error rate (WER) for English and character error rate (CER) for Mandarin, where lower values indicate better recognition performance.

As shown in Table~\ref{tab:asr_result}, despite its significantly smaller model size, Eureka-Audio-Instruct achieves competitive performance across all ASR benchmarks. Notably, on widely used English benchmarks such as LibriSpeech\cite{librispeech} and Fleurs\cite{fleurs}, Eureka-Audio attains lower error rates than several larger omni-modal and audio-centric models, including Qwen2.5-Omni and MiniCPM-o. These findings demonstrate that Eureka-Audio preserves robust speech recognition performance while attaining a desirable balance between predictive accuracy and computational efficiency.

\begin{table}[h!]
\centering
\small
\caption{ASR performance comparison. Using negative multirow prevents text clipping on colored backgrounds.}
\label{tab:asr_result}

\begin{tabularx}{\textwidth}{l l X c c}
\toprule
\textbf{Datasets} & \textbf{Type} & \textbf{Model} & \textbf{Size} & \textbf{WER/CER $\downarrow$} \\
\midrule

\multirow{11}{*}{\makecell[l]{\textbf{LibriSpeech} \\ test-clean \\ | test-other}}

& \multirow{5}{*}{\textit{Omni}}
  & Qwen3-Omni-Instruct & 30B-A3B & 1.60 | 2.93 \\
& & Ming-Lite-Omni-1.5 & 19B-A2.8B & 1.90 | 3.54 \\
& & MiniCPM-o & 9B & 2.01 | 4.87 \\
& & Qwen2.5-Omni-7B & 7B & 1.53 | 3.19 \\
& & Qwen2.5-Omni-3B & 3B & 1.68 | 3.90 \\

\cmidrule(lr){2-5}

& \multirow{4}{*}{\textit{Audio}}
  & Step-Audio-2-mini & 8B & 1.41 | 2.76 \\
& & Audio Flamingo 3 & 8B & 1.39 | 2.96 \\
& & Qwen2-Audio & 7B & 1.74 | 4.01 \\
& & Kimi-Audio-7B-Instruct & 7B & 1.33 | 2.57 \\

\cmidrule(lr){2-5}

\rowcolor{rowpink}
\cellcolor{white} &
  & Eureka-Audio-Base & 1.7B & 1.59 | 3.34 \\
\rowcolor{rowpink}
\cellcolor{white} & \multirow{-2}{*}{\textit{Ours}}
  & \textbf{Eureka-Audio-Instruct} & \textbf{1.7B} & \textbf{1.46 | 3.24} \\
\midrule

\multirow{11}{*}{\textbf{Fleurs-en}}
& \multirow{5}{*}{\textit{Omni}}
  & Qwen3-Omni-Instruct & 30B-A3B & 5.04 \\
& & Ming-Lite-Omni-1.5 & 19B-A2.8B & 5.82 \\
& & MiniCPM-o & 9B & 6.18 \\
& & Qwen2.5-Omni-7B & 7B & 5.49 \\
& & Qwen2.5-Omni-3B & 3B & 5.65 \\

\cmidrule(lr){2-5}

& \multirow{4}{*}{\textit{Audio}}
  & Step-Audio-2-mini & 8B & 4.51 \\
& & Audio Flamingo 3 & 8B & 6.30 \\
& & Qwen2-Audio & 7B & 6.92 \\
& & Kimi-Audio-7B-Instruct & 7B & 6.11 \\

\cmidrule(lr){2-5}

\rowcolor{rowpink}
\cellcolor{white} &
  & Eureka-Audio-Base & 1.7B & 5.73 \\
\rowcolor{rowpink}
\cellcolor{white} & \multirow{-2}{*}{\textit{Ours}}
  & \textbf{Eureka-Audio-Instruct} & \textbf{1.7B} & \textbf{5.39} \\
\midrule

\multirow{10}{*}{\makecell[l]{\textbf{AISHELL-2} \\ \textbf{ios}}}
& \multirow{5}{*}{\textit{Omni}}
  & Qwen3-Omni-Instruct & 30B-A3B & 2.63 \\
& & Ming-Lite-Omni-1.5 & 19B-A2.8B & 2.66 \\
& & MiniCPM-o & 9B & 3.42 \\
& & Qwen2.5-Omni-7B & 7B & 2.58 \\
& & Qwen2.5-Omni-3B & 3B & 2.77 \\

\cmidrule(lr){2-5}

& \multirow{3}{*}{\textit{Audio}}
  & Step-Audio-2-mini & 8B & 2.33 \\
& & Qwen2-Audio & 7B & 3.08 \\
& & Kimi-Audio-7B-Instruct & 7B & 2.80 \\

\cmidrule(lr){2-5}

\rowcolor{rowpink}
\cellcolor{white} &
  & Eureka-Audio-Base & 1.7B & 3.17 \\
\rowcolor{rowpink}
\cellcolor{white} & \multirow{-2}{*}{\textit{Ours}}
  & \textbf{Eureka-Audio-Instruct} & \textbf{1.7B} & \textbf{3.10} \\
\midrule

\multirow{10}{*}{\makecell[l]{\textbf{WenetSpeech} \\ test-meeting \\ | test-net}}
& \multirow{5}{*}{\textit{Omni}}
  & Qwen3-Omni-Instruct & 30B-A3B & 6.12 | 5.29 \\
& & Ming-Lite-Omni-1.5 & 19B-A2.8B & 5.96 | 6.26 \\
& & MiniCPM-o & 9B & 15.53 | 7.68 \\
& & Qwen2.5-Omni-7B & 7B & 8.43 | 7.10 \\
& & Qwen2.5-Omni-3B & 3B & 8.53 | 7.14 \\

\cmidrule(lr){2-5}

& \multirow{3}{*}{\textit{Audio}}
  & Step-Audio-2-mini & 8B & 5.43 | 5.50 \\
& & Qwen2-Audio & 7B & 8.40 | 8.00 \\
& & Kimi-Audio-7B-Instruct & 7B & 6.38 | 7.17 \\

\cmidrule(lr){2-5}

\rowcolor{rowpink}
\cellcolor{white} &
  & Eureka-Audio-Base & 1.7B & 10.37 | 8.63 \\
\rowcolor{rowpink}
\cellcolor{white} & \multirow{-2}{*}{\textit{Ours}}
  & \textbf{Eureka-Audio-Instruct} & \textbf{1.7B} & \textbf{9.14 | 7.55} \\
\bottomrule
\end{tabularx}
\end{table}

\subsection{Audio Understanding}

\begin{table}[h!]
\centering
\small
\caption{Performance comparison on audio understanding benchmarks. Following the standard layout with full-width spacing.}
\label{tab:audio_understanding}

\begin{tabularx}{\textwidth}{l l X c c}
\toprule
\textbf{Datasets} & \textbf{Type} & \textbf{Model} & \textbf{Size} & \textbf{Performance $\uparrow$} \\
\midrule

\multirow{11}{*}{\makecell[l]{\textbf{Knowledge} \\ MMSU \\ | OpenBookQA}}

& \multirow{5}{*}{\textit{Omni}}
  & Qwen3-Omni-Instruct & 30B-A3B & 77.00 | 92.31 \\
& & Ming-Lite-Omni-1.5 & 19B-A2.8B & 47.00 | 69.67 \\
& & MiniCPM-o & 9B & 54.55 | 79.12 \\
& & Qwen2.5-Omni-7B & 7B & 61.22 | 81.53 \\
& & Qwen2.5-Omni-3B & 3B & 53.41 | 77.36 \\

\cmidrule(lr){2-5}

& \multirow{4}{*}{\textit{Audio}}
  & Step-Audio-2-mini & 8B & 55.14 | 75.60 \\
& & Audio Flamingo 3 & 8B & 47.07 | 61.54 \\
& & Qwen2-Audio & 7B & 35.75 | 49.67 \\
& & Kimi-Audio-7B-Instruct & 7B & 61.26 | 84.18 \\

\cmidrule(lr){2-5}

\rowcolor{rowpink}
\cellcolor{white} &
  & Eureka-Audio-Base & 1.7B & 38.03 | 52.53 \\
\rowcolor{rowpink}
\cellcolor{white} & \multirow{-2}{*}{\textit{Ours}}
  & \textbf{Eureka-Audio-Instruct} & \textbf{1.7B} & \textbf{55.63 | 69.23} \\
\midrule

\multirow{10}{*}{\makecell[l]{\textbf{Safety} \\ AdvBench}}

& \multirow{5}{*}{\textit{Omni}}
  & Qwen3-Omni-Instruct & 30B-A3B & 99.61 \\
& & Ming-Lite-Omni-1.5 & 19B-A2.8B & 99.23 \\
& & MiniCPM-o & 9B & 95.76 \\
& & Qwen2.5-Omni-7B & 7B & 96.92 \\
& & Qwen2.5-Omni-3B & 3B & 89.80 \\

\cmidrule(lr){2-5}

& \multirow{4}{*}{\textit{Audio}}
  & Step-Audio-2-mini & 8B & 93.08 \\
& & Audio Flamingo 3 & 8B & 98.26 \\
& & Qwen2-Audio & 7B & 98.84 \\
& & Kimi-Audio-7B-Instruct & 7B & 100.00 \\

\cmidrule(lr){2-5}

\rowcolor{rowpink}
\cellcolor{white} & \textit{Ours}
  & \textbf{Eureka-Audio-Instruct} & \textbf{1.7B} & \textbf{99.81} \\
\midrule

\multirow{10}{*}{\makecell[l]{\textbf{Instruction} \\ IFEval}}

& \multirow{5}{*}{\textit{Omni}}
  & Qwen3-Omni-Instruct & 30B-A3B & 81.17 \\
& & Ming-Lite-Omni-1.5 & 19B-A2.8B & 53.68 \\
& & MiniCPM-o & 9B & 41.72 \\
& & Qwen2.5-Omni-7B & 7B & 39.84 \\
& & Qwen2.5-Omni-3B & 3B & 32.97 \\

\cmidrule(lr){2-5}

& \multirow{4}{*}{\textit{Audio}}
  & Step-Audio-2-mini & 8B & 43.54 \\
& & Audio Flamingo 3 & 8B & 32.27 \\
& & Qwen2-Audio & 7B & 26.24 \\
& & Kimi-Audio-7B-Instruct & 7B & 47.91 \\

\cmidrule(lr){2-5}

\rowcolor{rowpink}
\cellcolor{white} & \textit{Ours}
  & \textbf{Eureka-Audio-Instruct} & \textbf{1.7B} & \textbf{53.21} \\
\midrule

\multirow{12}{*}{\makecell[l]{\textbf{Paralinguistic} \\ MMAU$^\dagger$ \\ | MMAR}}

& \multirow{5}{*}{\textit{Omni}}
  & Qwen3-Omni-Instruct & 30B-A3B & 74.57 | 67.10 \\
& & Ming-Lite-Omni-1.5 & 19B-A2.8B & 63.52 | 45.40 \\
& & MiniCPM-o & 9B & 64.92 | 47.90 \\
& & Qwen2.5-Omni-7B & 7B & 66.23 | 49.60 \\
& & Qwen2.5-Omni-3B & 3B & 62.91 | 43.40 \\

\cmidrule(lr){2-5}

& \multirow{4}{*}{\textit{Audio}}
  & Step-Audio-2-mini & 8B & 71.96 | 61.57 \\
& & Audio Flamingo 3 & 8B & 74.77 | 61.00 \\
& & Qwen2-Audio & 7B & 59.80 | 37.90 \\
& & Kimi-Audio-7B-Instruct & 7B & 72.86 | 57.40 \\

\cmidrule(lr){2-5}

\rowcolor{rowpink}
\cellcolor{white} &
  & Eureka-Audio-Base & 1.7B & 63.42 | 46.80 \\
\rowcolor{rowpink}
\cellcolor{white} &
  & Eureka-Audio-Instruct w/o DataFlux & 1.7B & 66.93 | 50.70 \\
\rowcolor{rowpink}
\cellcolor{white} & \multirow{-3}{*}{\textit{Ours}}
  & \textbf{Eureka-Audio-Instruct w DataFlux} & \textbf{1.7B} & \textbf{74.67 | 56.20} \\
\bottomrule
\end{tabularx}
\end{table}

In this set of experiments, we cover both audio semantic understanding (e.g., audio-based factual reasoning and content comprehension) and audio paralinguistic understanding (e.g., emotion, environmental sounds, music, and reasoning based on fine-grained acoustic cues). We report results on a diverse set of benchmarks spanning multiple dimensions, including knowledge reasoning, instruction following, safety, and audio paralinguistic understanding. In particular, we adopt the MMAU~\cite{mmau} and MMAR~\cite{mmar} benchmarks to evaluate the model's ability to perform complex reasoning by going beyond literal speech content and jointly leveraging fine-grained acoustic cues.

As shown in Table~\ref{tab:audio_understanding}, despite containing only 1.7B parameters of LLM backbone, Eureka-Audio demonstrates competitive performance across a wide range of audio understanding benchmarks. Notably, on the MMAU benchmark, Eureka-Audio-Instruct outperforms several substantially larger models, including Qwen3-Omni-Instruct, achieves competitive or state-of-the-art-level performance. This result indicates that Eureka-Audio is capable of effectively modeling fine-grained audio semantic and paralinguistic information under limited model capacity, and exhibits strong reasoning ability on challenging audio understanding tasks.

Furthermore, after incorporating post-training data generated by DataFlux, the model achieves stable and consistent performance improvements on benchmarks related to audio paralinguistic understanding. This observation suggests that high-quality audio instruction data plays a critical role in stabilizing post-training and improving generalization, which is particularly important for lightweight models.

Overall, these results validate that Eureka-Audio achieves a favorable balance between efficiency and performance, enabling effective modeling of both audio semantic information and paralinguistic cues without relying on large-scale model parameters, and thereby supporting more general and fine-grained audio understanding capabilities.

\subsection{Dense Audio Captioning}
To further evaluate high-level audio understanding capabilities, we conduct experiments on dense audio captioning. This task requires the model to generate comprehensive and semantically rich descriptions that jointly capture speech content, acoustic events, and non-verbal cues, thereby preserving as much information from the original audio as possible.

To assess caption quality in a structured and quantifiable manner, we adopt a two-stage evaluation framework. First, the model generates a dense caption for the input audio. The generated caption is then concatenated with a downstream question and fed into a large language model (GPT-OSS-120B), which produces the final answer. Since the correctness of the answer critically depends on how faithfully the dense caption preserves the information contained in the original audio, this question answering conditioned on the generated caption pipeline serves as an indirect yet effective measure of dense caption quality. Models that generate more informative and structurally coherent captions are expected to achieve higher accuracy in the downstream reasoning stage.

As shown in Table~\ref{tab:dense_caption}, Eureka-Audio-Instruct outperforms Qwen3-Omni-Instruct on dense caption evaluation metrics, and its performance is comparable to the dedicated captioning model Qwen3-Omni-Captioner. These results indicate that, despite having only 1.7B parameters of LLM backbone, Eureka-Audio-Instruct is capable of producing information-rich and semantically faithful dense audio descriptions, demonstrating strong high-level audio understanding ability under a lightweight model setting.

\begin{table}[t]
\centering
\small
\caption{Dense audio caption evaluation results on MMAU and MMAR benchmarks.}
\label{tab:dense_caption}

\begin{tabularx}{0.9\textwidth}{l X c c}
\toprule
\textbf{Datasets} & \textbf{Model} & \textbf{Size} & \textbf{MMAU | MMAR $\uparrow$} \\
\midrule

\multirow{3}{*}{\makecell[l]{\textbf{Dense} \\ \textbf{Captioning}}}
& Qwen3-Omni-Captioner & 30B-A3B & 56.68 | 46.40 \\
& Qwen3-Omni-Instruct  & 30B-A3B & 48.24 | 36.90 \\

& \cellcolor{rowpink}\textbf{Eureka-Audio-Instruct (Ours)}
& \cellcolor{rowpink}\textbf{1.7B}
& \cellcolor{rowpink}\textbf{52.96 | 41.70} \\
\bottomrule
\end{tabularx}
\end{table}

\section{Conclusion}
In this work, we introduced Eureka-Audio, a lightweight 1.7B audio--language model designed to achieve strong audio semantic and paralinguistic understanding under strict parameter constraints. Through a unified end-to-end architecture integrating a Whisper-based audio encoder, a sparsely activated MoE adapter, and a compact language backbone, the model effectively aligns heterogeneous audio representations with limited capacity while mitigating cross-modal optimization conflicts.

Extensive evaluations across ASR, knowledge reasoning, safety, instruction following, and paralinguistic benchmarks demonstrate that Eureka-Audio consistently achieves competitive performance compared to substantially larger audio and omni-modal models. In particular, the model exhibits strong paralinguistic perception and reasoning ability, and benefits significantly from the proposed DataFlux pipeline, which provides structured, logically consistent audio instruction data to stabilize and enhance post-training.

Beyond accuracy, Eureka-Audio delivers up to a 3.7$\times$ speedup in decoding, highlighting a favorable efficiency and performance trade-off. These findings suggest that scaling model size is not the only path toward stronger audio intelligence; instead, careful architectural design and high-quality data synthesis can unlock competitive performance under constrained parameter budgets.

Looking ahead, we plan to extend Eureka-Audio beyond understanding toward unified audio generation and real-time interactive scenarios. Future work will explore lightweight and low-latency audio generation mechanisms, streaming architectures for real-time dialogue, and tighter integration between perception and generation within a single compact model. We believe such advancements will further bridge the gap between efficient modeling and practical audio intelligence deployment.

\clearpage
\bibliography{references}

@inproceedings{radford2023robust,
  title={Robust speech recognition via large-scale weak supervision},
  author={Radford, Alec and Kim, Jong Wook and Xu, Tao and Brockman, Greg and McLeavey, Christine and Sutskever, Ilya},
  booktitle={International conference on machine learning},
  pages={28492--28518},
  year={2023},
  organization={PMLR}
}

@misc{chu2023qwenaudioadvancinguniversalaudio,
  author        = {Yunfei Chu and Jin Xu and Xiaohuan Zhou and Qian Yang and Shiliang Zhang and Zhijie Yan and Chang Zhou and Jingren Zhou},
  title         = {Qwen{-}Audio: Advancing Universal Audio Understanding via Unified Large-Scale Audio{-}Language Models},
  year          = {2023},
  eprint        = {2311.07919},
  archiveprefix = {arXiv},
  primaryclass  = {eess.AS},
  url           = {https://arxiv.org/abs/2311.07919}
}

@article{Qwen2-audio,
  title={Qwen2-audio technical report},
  author={Chu, Yunfei and Xu, Jin and Yang, Qian and Wei, Haojie and Wei, Xipin and Guo, Zhifang and Leng, Yichong and Lv, Yuanjun and He, Jinzheng and Lin, Junyang and others},
  journal={arXiv preprint arXiv:2407.10759},
  year={2024}
}

@article{Kimi-Audio,
  title={Kimi-audio technical report},
  author={Ding, Ding and Ju, Zeqian and Leng, Yichong and Liu, Songxiang and Liu, Tong and Shang, Zeyu and Shen, Kai and Song, Wei and Tan, Xu and Tang, Heyi and others},
  journal={arXiv preprint arXiv:2504.18425},
  year={2025}
}

@article{Step-audio-2,
  title={Step-audio 2 technical report},
  author={Wu, Boyong and Yan, Chao and Hu, Chen and Yi, Cheng and Feng, Chengli and Tian, Fei and Shen, Feiyu and Yu, Gang and Zhang, Haoyang and Li, Jingbei and others},
  journal={arXiv preprint arXiv:2507.16632},
  year={2025}
}

@article{zeng2024glm,
  title={Glm-4-voice: Towards intelligent and human-like end-to-end spoken chatbot},
  author={Zeng, Aohan and Du, Zhengxiao and Liu, Mingdao and Wang, Kedong and Jiang, Shengmin and Zhao, Lei and Dong, Yuxiao and Tang, Jie},
  journal={arXiv preprint arXiv:2412.02612},
  year={2024}
}

@misc{fang2024llamaomniseamlessspeechinteraction,
  author        = {Qingkai Fang and Shoutao Guo and Yan Zhou and Zhengrui Ma and Shaolei Zhang and Yang Feng},
  title         = {LLaMA-Omni: Seamless Speech Interaction with Large Language Models},
  year          = {2024},
  eprint        = {2409.06666},
  archiveprefix = {arXiv},
  primaryclass  = {cs.CL},
  url           = {https://arxiv.org/abs/2409.06666}
}

@article{mentzer2023finite,
  title={Finite scalar quantization: Vq-vae made simple},
  author={Mentzer, Fabian and Minnen, David and Agustsson, Eirikur and Tschannen, Michael},
  journal={arXiv preprint arXiv:2309.15505},
  year={2023}
}

@article{ai2025ming,
  title={Ming-Omni: A Unified Multimodal Model for Perception and Generation},
  author={AI, Inclusion and Gong, Biao and Zou, Cheng and Zheng, Chuanyang and Zhou, Chunluan and Yan, Canxiang and Jin, Chunxiang and Shen, Chunjie and Zheng, Dandan and Wang, Fudong and others},
  journal={arXiv preprint arXiv:2506.09344},
  year={2025}
}

@article{qwen3-omni,
  title={Qwen3-Omni Technical Report},
  author={Jin Xu and Zhifang Guo and Hangrui Hu and Yunfei Chu and Xiong Wang and Jinzheng He and Yuxuan Wang and Xian Shi and Ting He and Xinfa Zhu and Yuanjun Lv and Yongqi Wang and Dake Guo and He Wang and Linhan Ma and Pei Zhang and Xinyu Zhang and Hongkun Hao and Zishan Guo and Baosong Yang and Bin Zhang and Ziyang Ma and Xipin Wei and Shuai Bai and Keqin Chen and Xuejing Liu and Peng Wang and Mingkun Yang and Dayiheng Liu and Xingzhang Ren and Bo Zheng and Rui Men and Fan Zhou and Bowen Yu and Jianxin Yang and Le Yu and Jingren Zhou and Junyang Lin},
  journal={arXiv preprint arXiv:2509.17765},
  year={2025}
}

@article{mmau,
  title={Mmau: A massive multi-task audio understanding and reasoning benchmark},
  author={Sakshi, S and Tyagi, Utkarsh and Kumar, Sonal and Seth, Ashish and Selvakumar, Ramaneswaran and Nieto, Oriol and Duraiswami, Ramani and Ghosh, Sreyan and Manocha, Dinesh},
  journal={arXiv preprint arXiv:2410.19168},
  year={2024}
}

@inproceedings{librispeech,
  title={Librispeech: an asr corpus based on public domain audio books},
  author={Panayotov, Vassil and Chen, Guoguo and Povey, Daniel and Khudanpur, Sanjeev},
  booktitle={2015 IEEE international conference on acoustics, speech and signal processing (ICASSP)},
  pages={5206--5210},
  year={2015},
  organization={IEEE}
}

@inproceedings{fleurs,
  title={Fleurs: Few-shot learning evaluation of universal representations of speech},
  author={Conneau, Alexis and Ma, Min and Khanuja, Simran and Zhang, Yu and Axelrod, Vera and Dalmia, Siddharth and Riesa, Jason and Rivera, Clara and Bapna, Ankur},
  booktitle={2022 IEEE Spoken Language Technology Workshop (SLT)},
  pages={798--805},
  year={2023},
  organization={IEEE}
}

@article{aishell-2,
  title={Aishell-2: Transforming mandarin asr research into industrial scale},
  author={Du, Jiayu and Na, Xingyu and Liu, Xuechen and Bu, Hui},
  journal={arXiv preprint arXiv:1808.10583},
  year={2018}
}

@article{glm4voice,
  title={Glm-4-voice: Towards intelligent and human-like end-to-end spoken chatbot},
  author={Zeng, Aohan and Du, Zhengxiao and Liu, Mingdao and Wang, Kedong and Jiang, Shengmin and Zhao, Lei and Dong, Yuxiao and Tang, Jie},
  journal={arXiv preprint arXiv:2412.02612},
  year={2024}
}

@misc{whisper,
  doi = {10.48550/ARXIV.2212.04356},
  url = {https://arxiv.org/abs/2212.04356},
  author = {Radford, Alec and Kim, Jong Wook and Xu, Tao and Brockman, Greg and McLeavey, Christine and Sutskever, Ilya},
  title = {Robust Speech Recognition via Large-Scale Weak Supervision},
  publisher = {arXiv},
  year = {2022},
  copyright = {arXiv.org perpetual, non-exclusive license}
}

@misc{moe-adapter,
      title={MoE Adapter for Large Audio Language Models: Sparsity, Disentanglement, and Gradient-Conflict-Free}, 
      author={Yishu Lei and Shuwei He and Jing Hu and Dan Zhang and Xianlong Luo and Danxiang Zhu and Shikun Feng and Rui Liu and Jingzhou He and Yu Sun and Hua Wu and Haifeng Wang},
      year={2026},
      eprint={2601.02967},
      archivePrefix={arXiv},
      primaryClass={cs.SD},
      url={https://arxiv.org/abs/2601.02967}, 
}

@article{Qwen3,
  title={Qwen3 technical report},
  author={Yang, An and Li, Anfeng and Yang, Baosong and Zhang, Beichen and Hui, Binyuan and Zheng, Bo and Yu, Bowen and Gao, Chang and Huang, Chengen and Lv, Chenxu and others},
  journal={arXiv preprint arXiv:2505.09388},
  year={2025}
}

@misc{gptoss,
      title={gpt-oss-120b \& gpt-oss-20b Model Card}, 
      author={OpenAI and : and Sandhini Agarwal and Lama Ahmad and Jason Ai and Sam Altman and Andy Applebaum and Edwin Arbus and Rahul K. Arora and Yu Bai and Bowen Baker and Haiming Bao and Boaz Barak and Ally Bennett and Tyler Bertao and Nivedita Brett and Eugene Brevdo and Greg Brockman and Sebastien Bubeck and Che Chang and Kai Chen and Mark Chen and Enoch Cheung and Aidan Clark and Dan Cook and Marat Dukhan and Casey Dvorak and Kevin Fives and Vlad Fomenko and Timur Garipov and Kristian Georgiev and Mia Glaese and Tarun Gogineni and Adam Goucher and Lukas Gross and Katia Gil Guzman and John Hallman and Jackie Hehir and Johannes Heidecke and Alec Helyar and Haitang Hu and Romain Huet and Jacob Huh and Saachi Jain and Zach Johnson and Chris Koch and Irina Kofman and Dominik Kundel and Jason Kwon and Volodymyr Kyrylov and Elaine Ya Le and Guillaume Leclerc and James Park Lennon and Scott Lessans and Mario Lezcano-Casado and Yuanzhi Li and Zhuohan Li and Ji Lin and Jordan Liss and Lily and Liu and Jiancheng Liu and Kevin Lu and Chris Lu and Zoran Martinovic and Lindsay McCallum and Josh McGrath and Scott McKinney and Aidan McLaughlin and Song Mei and Steve Mostovoy and Tong Mu and Gideon Myles and Alexander Neitz and Alex Nichol and Jakub Pachocki and Alex Paino and Dana Palmie and Ashley Pantuliano and Giambattista Parascandolo and Jongsoo Park and Leher Pathak and Carolina Paz and Ludovic Peran and Dmitry Pimenov and Michelle Pokrass and Elizabeth Proehl and Huida Qiu and Gaby Raila and Filippo Raso and Hongyu Ren and Kimmy Richardson and David Robinson and Bob Rotsted and Hadi Salman and Suvansh Sanjeev and Max Schwarzer and D. Sculley and Harshit Sikchi and Kendal Simon and Karan Singhal and Yang Song and Dane Stuckey and Zhiqing Sun and Philippe Tillet and Sam Toizer and Foivos Tsimpourlas and Nikhil Vyas and Eric Wallace and Xin Wang and Miles Wang and Olivia Watkins and Kevin Weil and Amy Wendling and Kevin Whinnery and Cedric Whitney and Hannah Wong and Lin Yang and Yu Yang and Michihiro Yasunaga and Kristen Ying and Wojciech Zaremba and Wenting Zhan and Cyril Zhang and Brian Zhang and Eddie Zhang and Shengjia Zhao},
      year={2025},
      eprint={2508.10925},
      archivePrefix={arXiv},
      primaryClass={cs.CL},
      url={https://arxiv.org/abs/2508.10925}, 
}

@misc{Step-Audio-R1,
      title={Step-Audio-R1 Technical Report}, 
      author={Fei Tian and Xiangyu Tony Zhang and Yuxin Zhang and Haoyang Zhang and Yuxin Li and Daijiao Liu and Yayue Deng and Donghang Wu and Jun Chen and Liang Zhao and Chengyuan Yao and Hexin Liu and Eng Siong Chng and Xuerui Yang and Xiangyu Zhang and Daxin Jiang and Gang Yu},
      year={2025},
      eprint={2511.15848},
      archivePrefix={arXiv},
      primaryClass={cs.AI},
      url={https://arxiv.org/abs/2511.15848}, 
}

@inproceedings{wenetspeech,
  title={Wenetspeech: A 10000+ hours multi-domain mandarin corpus for speech recognition},
  author={Zhang, Binbin and Lv, Hang and Guo, Pengcheng and Shao, Qijie and Yang, Chao and Xie, Lei and Xu, Xin and Bu, Hui and Chen, Xiaoyu and Zeng, Chenchen and others},
  booktitle={ICASSP 2022-2022 IEEE International Conference on Acoustics, Speech and Signal Processing (ICASSP)},
  pages={6182--6186},
  year={2022},
  organization={IEEE}
}

@misc{mmar,
      title={MMAR: A Challenging Benchmark for Deep Reasoning in Speech, Audio, Music, and Their Mix}, 
      author={Ziyang Ma and Yinghao Ma and Yanqiao Zhu and Chen Yang and Yi-Wen Chao and Ruiyang Xu and Wenxi Chen and Yuanzhe Chen and Zhuo Chen and Jian Cong and Kai Li and Keliang Li and Siyou Li and Xinfeng Li and Xiquan Li and Zheng Lian and Yuzhe Liang and Minghao Liu and Zhikang Niu and Tianrui Wang and Yuping Wang and Yuxuan Wang and Yihao Wu and Guanrou Yang and Jianwei Yu and Ruibin Yuan and Zhisheng Zheng and Ziya Zhou and Haina Zhu and Wei Xue and Emmanouil Benetos and Kai Yu and Eng-Siong Chng and Xie Chen},
      year={2025},
      eprint={2505.13032},
      archivePrefix={arXiv},
      primaryClass={cs.SD},
      url={https://arxiv.org/abs/2505.13032}, 
}

@misc{heittola2022tau,
  author    = {Heittola, T. and Mesaros, A. and Virtanen, T. and Heittola, T. and Laakso, H. and Bejarano Rodriguez, R.},
  title     = {TAU Urban Acoustic Scenes 2022 Mobile, Development dataset},
  year      = {2022},
  month     = {03},
  day       = {8},
  howpublished = {Zenodo},
  doi       = {10.5281/zenodo.6337421},
}

@inproceedings{jeong2022cochlscene,
  title={Cochlscene: Acquisition of acoustic scene data using crowdsourcing},
  author={Jeong, Il-Young and Park, Jeongsoo},
  booktitle={2022 Asia-Pacific Signal and Information Processing Association Annual Summit and Conference (APSIPA ASC)},
  pages={17--21},
  year={2022},
  organization={IEEE}
}

@inproceedings{Mesaros2016_EUSIPCO,
    author = "Mesaros, Annamaria and Heittola, Toni and Virtanen, Tuomas",
    title = "{TUT} Database for Acoustic Scene Classification and Sound Event Detection",
    year = "2016",
    address = "Budapest, Hungary",
    booktitle = "24th European Signal Processing Conference 2016 (EUSIPCO 2016)"
}

@article{zhou2022emotional,
title={Emotional voice conversion: Theory, databases and ESD},
author={Zhou, Kun and Sisman, Berrak and Liu, Rui and Li, Haizhou},
journal={Speech Communication},
volume={137},
pages={1--18},
year={2022},
publisher={Elsevier}
}

@article{tripathi2018multi,
  title={Multi-modal emotion recognition on iemocap dataset using deep learning},
  author={Tripathi, Samarth and Tripathi, Sarthak and Beigi, Homayoon},
  journal={arXiv preprint arXiv:1804.05788},
  year={2018}
}

@article{poria2018meld,
  title={Meld: A multimodal multi-party dataset for emotion recognition in conversations},
  author={Poria, Soujanya and Hazarika, Devamanyu and Majumder, Navonil and Naik, Gautam and Cambria, Erik and Mihalcea, Rada},
  journal={arXiv preprint arXiv:1810.02508},
  year={2018}
}

@article{livingstone2018ryerson,
  title={The Ryerson Audio-Visual Database of Emotional Speech and Song (RAVDESS): A dynamic, multimodal set of facial and vocal expressions in North American English},
  author={Livingstone, Steven R and Russo, Frank A},
  journal={PloS one},
  volume={13},
  number={5},
  pages={e0196391},
  year={2018},
  publisher={Public Library of Science San Francisco, CA USA}
}

@misc{hajiali2025tamingdatatransformersaudio,
      title={Taming Data and Transformers for Audio Generation}, 
      author={Moayed Haji-Ali and Willi Menapace and Aliaksandr Siarohin and Guha Balakrishnan and Vicente Ordonez},
      year={2025},
      eprint={2406.19388},
      archivePrefix={arXiv},
      primaryClass={cs.SD},
      url={https://arxiv.org/abs/2406.19388}, 
}

@inproceedings{drossos2020clotho,
  title={Clotho: An audio captioning dataset},
  author={Drossos, Konstantinos and Lipping, Samuel and Virtanen, Tuomas},
  booktitle={ICASSP 2020-2020 IEEE International Conference on Acoustics, Speech and Signal Processing (ICASSP)},
  pages={736--740},
  year={2020},
  organization={IEEE}
}

@misc{hung2024tangofluxsuperfastfaithful,
      title={TangoFlux: Super Fast and Faithful Text to Audio Generation with Flow Matching and Clap-Ranked Preference Optimization}, 
      author={Chia-Yu Hung and Navonil Majumder and Zhifeng Kong and Ambuj Mehrish and Rafael Valle and Bryan Catanzaro and Soujanya Poria},
      year={2024},
      eprint={2412.21037},
      archivePrefix={arXiv},
      primaryClass={cs.SD},
      url={https://arxiv.org/abs/2412.21037}, 
}

@inproceedings{martin2021ground,
  title={What is the ground truth? reliability of multi-annotator data for audio tagging},
  author={Mart{\'\i}n-Morat{\'o}, Irene and Mesaros, Annamaria},
  booktitle={2021 29th European Signal Processing Conference (EUSIPCO)},
  pages={76--80},
  year={2021},
  organization={IEEE}
}

@article{koepke2022audio,
  title={Audio retrieval with natural language queries: A benchmark study},
  author={Koepke, A Sophia and Oncescu, Andreea-Maria and Henriques, Jo{\~a}o F and Akata, Zeynep and Albanie, Samuel},
  journal={IEEE Transactions on Multimedia},
  volume={25},
  pages={2675--2685},
  year={2022},
  publisher={IEEE}
}

@misc{primus2025tacostemporallyalignedaudiocaptions,
      title={TACOS: Temporally-aligned Audio CaptiOnS for Language-Audio Pretraining}, 
      author={Paul Primus and Florian Schmid and Gerhard Widmer},
      year={2025},
      eprint={2505.07609},
      archivePrefix={arXiv},
      primaryClass={eess.AS},
      url={https://arxiv.org/abs/2505.07609}, 
}

@inproceedings{chen2020vggsound,
  title={Vggsound: A large-scale audio-visual dataset},
  author={Chen, Honglie and Xie, Weidi and Vedaldi, Andrea and Zisserman, Andrew},
  booktitle={ICASSP 2020-2020 IEEE International Conference on Acoustics, Speech and Signal Processing (ICASSP)},
  pages={721--725},
  year={2020},
  organization={IEEE}
}

@article{mei2024wavcaps,
  title={Wavcaps: A chatgpt-assisted weakly-labelled audio captioning dataset for audio-language multimodal research},
  author={Mei, Xinhao and Meng, Chutong and Liu, Haohe and Kong, Qiuqiang and Ko, Tom and Zhao, Chengqi and Plumbley, Mark D and Zou, Yuexian and Wang, Wenwu},
  journal={IEEE/ACM Transactions on Audio, Speech, and Language Processing},
  year={2024},
  publisher={IEEE}
}

@inproceedings{piczak2015dataset,
  title = {{ESC}: {Dataset} for {Environmental Sound Classification}},
  author = {Piczak, Karol J.},
  booktitle = {Proceedings of the 23rd {Annual ACM Conference} on {Multimedia}},
  date = {2015-10-13},
  url = {http://dl.acm.org/citation.cfm?doid=2733373.2806390},
  doi = {10.1145/2733373.2806390},
  location = {{Brisbane, Australia}},
  isbn = {978-1-4503-3459-4},
  publisher = {{ACM Press}},
  pages = {1015--1018}
}

@inproceedings{kim2019audiocaps,
  title={Audiocaps: Generating captions for audios in the wild},
  author={Kim, Chris Dongjoo and Kim, Byeongchang and Lee, Hyunmin and Kim, Gunhee},
  booktitle={Proceedings of the 2019 Conference of the North American Chapter of the Association for Computational Linguistics: Human Language Technologies, Volume 1 (Long and Short Papers)},
  pages={119--132},
  year={2019}
}

@article{rashid2023nonspeech7k,
  title={Nonspeech7k dataset: Classification and analysis of human non-speech sound},
  author={Rashid, Muhammad Mamunur and Li, Guiqing and Du, Chengrui},
  journal={IET Signal Processing},
  volume={17},
  number={6},
  pages={e12233},
  year={2023},
  publisher={Wiley Online Library}
}

@inproceedings{salamon2014dataset,
  title={A dataset and taxonomy for urban sound research},
  author={Salamon, Justin and Jacoby, Christopher and Bello, Juan Pablo},
  booktitle={Proceedings of the 22nd ACM international conference on Multimedia},
  pages={1041--1044},
  year={2014}
}

@INPROCEEDINGS{gong_vocalsound,
  author={Gong, Yuan and Yu, Jin and Glass, James},
  booktitle={ICASSP 2022 - 2022 IEEE International Conference on Acoustics, Speech and Signal Processing (ICASSP)}, 
  title={Vocalsound: A Dataset for Improving Human Vocal Sounds Recognition}, 
  year={2022},
  pages={151-155},
  doi={10.1109/ICASSP43922.2022.9746828}}

@misc{chen2025fusionaudio12mfinegrainedaudiocaptioning,
      title={FusionAudio-1.2M: Towards Fine-grained Audio Captioning with Multimodal Contextual Fusion}, 
      author={Shunian Chen and Xinyuan Xie and Zheshu Chen and Liyan Zhao and Owen Lee and Zhan Su and Qilin Sun and Benyou Wang},
      year={2025},
      eprint={2506.01111},
      archivePrefix={arXiv},
      primaryClass={cs.SD},
      url={https://arxiv.org/abs/2506.01111}, 
}

@article{aqa2025multi,
  title={Multi-Domain Audio Question Answering Toward Acoustic Content Reasoning in The DCASE 2025 Challenge},
  author={Yang, Chao-Han Huck and Ghosh, Sreyan and Wang, Qing and Kim, Jaeyeon and Hong, Hengyi and Kumar, Sonal and Zhong, Guirui and Kong, Zhifeng and Sakshi, S and Lokegaonkar, Vaibhavi and Nieto, Oriol and Duraiswami, Ramani and Manocha, Dinesh and Kim, Gunhee and Du, Jun and Valle, Rafael and Catanzaro, Bryan},
  journal={arXiv preprint arXiv:2505.07365},
  year={2025},
  url={https://arxiv.org/abs/2505.07365}
}

@article{defferrard2016fma,
  title={FMA: A dataset for music analysis},
  author={Defferrard, Micha{\"e}l and Benzi, Kirell and Vandergheynst, Pierre and Bresson, Xavier},
  journal={arXiv preprint arXiv:1612.01840},
  year={2016}
}

@article{doh2023lp,
  title={Lp-musiccaps: Llm-based pseudo music captioning},
  author={Doh, SeungHeon and Choi, Keunwoo and Lee, Jongpil and Nam, Juhan},
  journal={arXiv preprint arXiv:2307.16372},
  year={2023}
}

@article{agostinelli2023musiclm,
  title={Musiclm: Generating music from text},
  author={Agostinelli, Andrea and Denk, Timo I and Borsos, Zal{\'a}n and Engel, Jesse and Verzetti, Mauro and Caillon, Antoine and Huang, Qingqing and Jansen, Aren and Roberts, Adam and Tagliasacchi, Marco and others},
  journal={arXiv preprint arXiv:2301.11325},
  year={2023}
}

@article{melechovsky2023mustango,
  title={Mustango: Toward controllable text-to-music generation},
  author={Melechovsky, Jan and Guo, Zixun and Ghosal, Deepanway and Majumder, Navonil and Herremans, Dorien and Poria, Soujanya},
  journal={arXiv preprint arXiv:2311.08355},
  year={2023}
}

@misc{manco2023songdescriberdatasetcorpus,
      title={The Song Describer Dataset: a Corpus of Audio Captions for Music-and-Language Evaluation}, 
      author={Ilaria Manco and Benno Weck and SeungHeon Doh and Minz Won and Yixiao Zhang and Dmitry Bogdanov and Yusong Wu and Ke Chen and Philip Tovstogan and Emmanouil Benetos and Elio Quinton and György Fazekas and Juhan Nam},
      year={2023},
      eprint={2311.10057},
      archivePrefix={arXiv},
      primaryClass={cs.SD},
      url={https://arxiv.org/abs/2311.10057}, 
}

@misc{xie2024picoaudioenablingprecisetimestamp,
      title={PicoAudio: Enabling Precise Timestamp and Frequency Controllability of Audio Events in Text-to-audio Generation}, 
      author={Zeyu Xie and Xuenan Xu and Zhizheng Wu and Mengyue Wu},
      year={2024},
      eprint={2407.02869},
      archivePrefix={arXiv},
      primaryClass={cs.SD},
      url={https://arxiv.org/abs/2407.02869}, 
}

@misc{diwan2025scalingrichstylepromptedtexttospeech,
      title={Scaling Rich Style-Prompted Text-to-Speech Datasets}, 
      author={Anuj Diwan and Zhisheng Zheng and David Harwath and Eunsol Choi},
      year={2025},
      eprint={2503.04713},
      archivePrefix={arXiv},
      primaryClass={eess.AS},
      url={https://arxiv.org/abs/2503.04713}, 
}

@article{he2025emilia,
  title={Emilia: A Large-Scale, Extensive, Multilingual, and Diverse Dataset for Speech Generation},
  author={He, Haorui and Shang, Zengqiang and Wang, Chaoren and Li, Xuyuan and Gu, Yicheng and Hua, Hua and Liu, Liwei and Yang, Chen and Li, Jiaqi and Shi, Peiyang and others},
  journal={arXiv preprint arXiv:2501.15907},
  year={2025}
}

@inproceedings{bu2017aishell,
  title={Aishell-1: An open-source mandarin speech corpus and a speech recognition baseline},
  author={Bu, Hui and Du, Jiayu and Na, Xingyu and Wu, Bengu and Zheng, Hao},
  booktitle={2017 20th conference of the oriental chapter of the international coordinating committee on speech databases and speech I/O systems and assessment (O-COCOSDA)},
  pages={1--5},
  year={2017},
  organization={IEEE}
}

@article{du2018aishell,
  title={Aishell-2: Transforming mandarin asr research into industrial scale},
  author={Du, Jiayu and Na, Xingyu and Liu, Xuechen and Bu, Hui},
  journal={arXiv preprint arXiv:1808.10583},
  year={2018}
}

@article{shi2020aishell,
  title={Aishell-3: A multi-speaker mandarin tts corpus and the baselines},
  author={Shi, Yao and Bu, Hui and Xu, Xin and Zhang, Shaoji and Li, Ming},
  journal={arXiv preprint arXiv:2010.11567},
  year={2020}
}

@article{chen2021gigaspeech,
  title={Gigaspeech: An evolving, multi-domain asr corpus with 10,000 hours of transcribed audio},
  author={Chen, Guoguo and Chai, Shuzhou and Wang, Guanbo and Du, Jiayu and Zhang, Wei-Qiang and Weng, Chao and Su, Dan and Povey, Daniel and Trmal, Jan and Zhang, Junbo and others},
  journal={arXiv preprint arXiv:2106.06909},
  year={2021}
}

@article{zen2019libritts,
  title={Libritts: A corpus derived from librispeech for text-to-speech},
  author={Zen, Heiga and Dang, Viet and Clark, Rob and Zhang, Yu and Weiss, Ron J and Jia, Ye and Chen, Zhifeng and Wu, Yonghui},
  journal={arXiv preprint arXiv:1904.02882},
  year={2019}
}

@inproceedings{kang2024libriheavy,
  title={Libriheavy: A 50,000 hours ASR corpus with punctuation casing and context},
  author={Kang, Wei and Yang, Xiaoyu and Yao, Zengwei and Kuang, Fangjun and Yang, Yifan and Guo, Liyong and Lin, Long and Povey, Daniel},
  booktitle={ICASSP 2024-2024 IEEE International Conference on Acoustics, Speech and Signal Processing (ICASSP)},
  pages={10991--10995},
  year={2024},
  organization={IEEE}
}

@inproceedings{panayotov2015librispeech,
  title={Librispeech: an asr corpus based on public domain audio books},
  author={Panayotov, Vassil and Chen, Guoguo and Povey, Daniel and Khudanpur, Sanjeev},
  booktitle={2015 IEEE international conference on acoustics, speech and signal processing (ICASSP)},
  pages={5206--5210},
  year={2015},
  organization={IEEE}
}

@inproceedings{tang2021kespeech,
  title={Kespeech: An open source speech dataset of mandarin and its eight subdialects},
  author={Tang, Zhiyuan and Wang, Dong and Xu, Yanguang and Sun, Jianwei and Lei, Xiaoning and Zhao, Shuaijiang and Wen, Cheng and Tan, Xingjun and Xie, Chuandong and Zhou, Shuran and others},
  booktitle={Thirty-fifth Conference on Neural Information Processing Systems Datasets and Benchmarks Track (Round 2)},
  year={2021}
}

@article{yang2022open,
  title={Open source magicdata-ramc: A rich annotated mandarin conversational (ramc) speech dataset},
  author={Yang, Zehui and Chen, Yifan and Luo, Lei and Yang, Runyan and Ye, Lingxuan and Cheng, Gaofeng and Xu, Ji and Jin, Yaohui and Zhang, Qingqing and Zhang, Pengyuan and others},
  journal={arXiv preprint arXiv:2203.16844},
  year={2022}
}

@inproceedings{zhang2022wenetspeech,
  title={Wenetspeech: A 10000+ hours multi-domain mandarin corpus for speech recognition},
  author={Zhang, Binbin and Lv, Hang and Guo, Pengcheng and Shao, Qijie and Yang, Chao and Xie, Lei and Xu, Xin and Bu, Hui and Chen, Xiaoyu and Zeng, Chenchen and others},
  booktitle={ICASSP 2022-2022 IEEE International Conference on Acoustics, Speech and Signal Processing (ICASSP)},
  pages={6182--6186},
  year={2022},
  organization={IEEE}
}

@article{ma2024wenetspeech4tts,
  title={Wenetspeech4tts: A 12,800-hour mandarin tts corpus for large speech generation model benchmark},
  author={Ma, Linhan and Guo, Dake and Song, Kun and Jiang, Yuepeng and Wang, Shuai and Xue, Liumeng and Xu, Weiming and Zhao, Huan and Zhang, Binbin and Xie, Lei},
  journal={arXiv preprint arXiv:2406.05763},
  year={2024}
}

@article{xu2024magpie,
  title={Magpie: Alignment data synthesis from scratch by prompting aligned llms with nothing},
  author={Xu, Zhangchen and Jiang, Fengqing and Niu, Luyao and Deng, Yuntian and Poovendran, Radha and Choi, Yejin and Lin, Bill Yuchen},
  journal={arXiv preprint arXiv:2406.08464},
  year={2024}
}

@article{Pratap2020MLSAL,
  title={MLS: A Large-Scale Multilingual Dataset for Speech Research},
  author={Vineel Pratap and Qiantong Xu and Anuroop Sriram and Gabriel Synnaeve and Ronan Collobert},
  journal={ArXiv},
  year={2020},
  volume={abs/2012.03411}
}

@inproceedings{conneau2023fleurs,
  title={Fleurs: Few-shot learning evaluation of universal representations of speech},
  author={Conneau, Alexis and Ma, Min and Khanuja, Simran and Zhang, Yu and Axelrod, Vera and Dalmia, Siddharth and Riesa, Jason and Rivera, Clara and Bapna, Ankur},
  booktitle={2022 IEEE Spoken Language Technology Workshop (SLT)},
  pages={798--805},
  year={2023},
  organization={IEEE}
}

@inproceedings{hershey2021benefit,
  title={The benefit of temporally-strong labels in audio event classification},
  author={Hershey, Shawn and Ellis, Daniel PW and Fonseca, Eduardo and Jansen, Aren and Liu, Caroline and Moore, R Channing and Plakal, Manoj},
  booktitle={ICASSP 2021-2021 IEEE International Conference on Acoustics, Speech and Signal Processing (ICASSP)},
  pages={366--370},
  year={2021},
  organization={IEEE}
}

@article{bakhturina2021hi,
  title={{Hi-Fi Multi-Speaker English TTS Dataset}},
  author={Bakhturina, Evelina and Lavrukhin, Vitaly and Ginsburg, Boris and Zhang, Yang},
  journal={arXiv preprint arXiv:2104.01497},
  year={2021}
}

@misc{wang2024globehighqualityenglishcorpus,
      title={GLOBE: A High-quality English Corpus with Global Accents for Zero-shot Speaker Adaptive Text-to-Speech}, 
      author={Wenbin Wang and Yang Song and Sanjay Jha},
      year={2024},
      eprint={2406.14875},
      archivePrefix={arXiv},
      primaryClass={cs.SD},
      url={https://arxiv.org/abs/2406.14875}, 
}
\bibliographystyle{unsrt}

\clearpage
\section{Appendix}
\subsection{Open-Source Audio Datasets for Pretraining}
\label{app:opensource_audio_for_pretraining}

In Table~\ref{tab:pretrain_data}, we list all audio caption data used in the pretrain stage.
\begin{table}[H]
    \centering
    \caption{List of datasets used for audio caption in Pretraining stage.}
    \label{tab:pretrain_data}
    \setlength{\tabcolsep}{0pt}
    \begin{tabular*}{\textwidth}{@{\extracolsep{\fill}}lc lc}
    \toprule
    \textbf{Dataset} & \textbf{Audio Length (Hours)} & \textbf{Dataset} & \textbf{Audio Length (Hours)} \\
    \midrule
    CochlScene~\citep{jeong2022cochlscene} & 169 & TACOS~\citep{primus2025tacostemporallyalignedaudiocaptions} & 76 \\
    TUT2016~\citep{Mesaros2016_EUSIPCO} & 10 & VGGSound~\citep{chen2020vggsound} & 513 \\
    TUT2017~\citep{Mesaros2016_EUSIPCO} & 13 & WavCaps~\citep{mei2024wavcaps} & 3,793 \\
    TAU2022~\citep{heittola2022tau} & 67 & ESC50~\citep{piczak2015dataset} & 1 \\
    ESD~\citep{zhou2022emotional} & 29 & Nonspeech7k~\citep{rashid2023nonspeech7k} & 6 \\
    IEMOCAP~\citep{tripathi2018multi} & 10 & UrbanSound8K~\citep{salamon2014dataset} & 9 \\
    MELD~\citep{poria2018meld} & 9  & VocalSound~\citep{gong_vocalsound} & 19 \\
    RAVDESS~\citep{livingstone2018ryerson} & 3 & FusionAudio~\citep{chen2025fusionaudio12mfinegrainedaudiocaptioning} & 16,646 \\
    AudioCaps~\citep{kim2019audiocaps} & 137 & DCASEAudioQA~\citep{aqa2025multi} & 57 \\
    AudioSet Strong~\citep{hershey2021benefit} & 352 & FMA~\cite{defferrard2016fma} & 860 \\
    AutoReCap\citep{hajiali2025tamingdatatransformersaudio} & 1,235,000 & LP-MusicCaps\textsubscript{MC}~\cite{doh2023lp} & 7 \\
    Clotho~\citep{drossos2020clotho} & 17 & MusicCaps~\cite{agostinelli2023musiclm} & 7 \\
    Clotho-v2~\citep{drossos2020clotho} & 26 & MusicBench~\cite{melechovsky2023mustango} & 115 \\
    CRPO~\citep{hung2024tangofluxsuperfastfaithful} & 277 & SDD~\citep{manco2023songdescriberdatasetcorpus} & 36 \\
    MACS~\citep{martin2021ground} & 10 & PicoAudio~\citep{xie2024picoaudioenablingprecisetimestamp} & 12 \\
    SoundDescs~\citep{koepke2022audio} & 1,060 & ParaSpeechCaps~\citep{diwan2025scalingrichstylepromptedtexttospeech} & 2,769 \\
    \bottomrule
    \end{tabular*}
\end{table}

\subsection{Open-Source Audio Datasets for Post-training}
\label{app:opensource_audio_for_posttraining}

In Table~\ref{tab:sft_data}, we list all ASR data used in the post-train stage.
\begin{table}[H]
    \centering
    \caption{List of datasets used for ASR in Post-training stage.}
    \label{tab:sft_data}
    \setlength{\tabcolsep}{0pt}
    \begin{tabular*}{\textwidth}{@{\extracolsep{\fill}}lcc}
    \toprule
    \textbf{Dataset} & \textbf{Language} & \textbf{Audio Length (Hours)} \\
    \midrule
    Emilia~\citep{he2025emilia} & Multi & 98,305 \\
    AISHELL-1~\citep{bu2017aishell} & zh & 155 \\
    AISHELL-2~\citep{du2018aishell} & zh & 1,036 \\
    AISHELL-3~\citep{shi2020aishell} & zh & 65 \\
    ESD~\citep{zhou2022emotional} & zh, en & 29 \\
    Gigaspeech~\citep{chen2021gigaspeech} & en & 10,288 \\
    Hi-Fi TTS~\citep{bakhturina2021hi} & en & 291 \\
    GLOBE~\citep{wang2024globehighqualityenglishcorpus} & en & 535 \\
    LibriTTS~\citep{zen2019libritts} & en & 568 \\
    Libriheavy~\citep{kang2024libriheavy} & en & 51,448 \\
    LibriSpeech~\citep{panayotov2015librispeech} & en & 960 \\
    KeSpeech~\citep{tang2021kespeech} & zh & 1,428 \\
    Magicdata~\citep{yang2022open} & zh & 747 \\
    WenetSpeech~\citep{zhang2022wenetspeech} & zh & 10,518 \\
    WenetSpeech4TTS~\citep{ma2024wenetspeech4tts} & zh & 12,085 \\
    zhvoice\textsuperscript{1} & zh & 901 \\
    Magpie~\citep{xu2024magpie} & en & 306 \\
    MLS~\citep{Pratap2020MLSAL} & Multi & 45,042 \\
    Fleurs~\citep{conneau2023fleurs} & Multi & 17 \\
    \bottomrule
    \end{tabular*}
\end{table}

\subsection{Prompts for DataFlux}
\label{app:prompts_for_dataflux}
In this section, we present the prompt templates used in the DataFlux pipeline, including those for query generation and answer verification.

\begin{tcolorbox}[
    enhanced,
    colback=white,
    colbacklower=gray!20,
    colframe=black,
    boxrule=2pt,
    sharp corners,
    width=\textwidth,
    title={Query\&Choices Generation},
    colbacktitle=black,
    coltitle=white,
    fonttitle=\bfseries\sffamily\large,
    halign title=center,
    breakable,
    listing only,
    listing options={
        basicstyle=\ttfamily\footnotesize,
        columns=fullflexible,
        breaklines=false,
        literate={\textbackslash n}{{\textbackslash n}}2
    }
]

\# ROLE\\
You are an expert Audio QA Dataset Curator. Your goal is to create a high-quality "Listening Comprehension" test based on ground-truth audio descriptions.\\

\# TASK SPECIFICATION\\
**STEP 1: VERIFICATION**\\
Analyze the provided `Audio Caption` against the `Category Hierarchy` (L0-L3).\\
- Determine if the event described in the caption logically falls under the specific Sub-Task/Category provided.\\
- **Mismatch Example:** Category is "Animal Sounds > Dog" but Caption describes "A car engine revving". (Result: `is\_success`: false)\\
- **Match Example:** Category is "Animal Sounds > Dog" and Caption describes "A large dog barking aggressively". (Result: `is\_success`: true)\\

**STEP 2: GENERATION (Only if Match)**\\
If the caption is consistent, generate **5 distinct Multiple Choice Questions (MCQs)** that a user would answer by **LISTENING to the audio**.\\
- **Format:** Each item must be a single string containing the **Question** followed by **4 Options (A, B, C, D)** separated by newlines (`\textbackslash n`).\\
- **Distractors:** Wrong options must be plausible audio events relevant to the context, not obviously ridiculous.\\
- **Randomness:** Randomize the position of the correct option.\\
- **Content:** Questions must target the acoustic properties, events, objects, or environment audible in the recording.\\

\# CRITICAL CONSTRAINTS (MUST FOLLOW)\\
1. **AUDIO-CENTRIC PERSPECTIVE:** Questions must be phrased as if the user is currently hearing the sound. Use phrases like "What sound is heard...", "Identify the...", "What action causes the noise...", etc.\\
2. **FORBIDDEN PHRASES:** You are strictly **FORBIDDEN** from using words that refer to the text, such as "According to the caption", "What does the description say", etc.\\
3. **NO ANSWERS REVEALED:** Do NOT include the correct answer key or mark the correct option in the output string.\\
4. **SINGLE STRING FORMAT:** The question and options must be combined into one string using `\textbackslash n`.\\

\# HIERARCHICAL CONTEXT\\
\{\{HIERARCHY\_BLOCK\}\}\\

\# INPUT DATA\\
- **Audio Caption:** "\{\{AUDIO\_CAPTION\}\}"\\

\# FEW-SHOT EXAMPLES (REFERENCE ONLY)\\
\{\{FEW\_SHOT\_CONTENT\}\}\\

\# OUTPUT FORMAT\\
You must output a single valid JSON object. Do not include markdown formatting (like ```json).\\

**Structure for Valid Match:**\\
\{\\
  "is\_success": true,\\
  "questions": [\\
    "What is the primary sound heard?\textbackslash nA. A dog barking\textbackslash nB. A car honking\textbackslash nC. People talking\textbackslash nD. Birds chirping",\\
    "Identify the background noise.\textbackslash nA. Wind blowing\textbackslash nB. Rain falling\textbackslash nC. Traffic noise\textbackslash nD. Silence",\\
    "How would you describe the tone of the voice?\textbackslash nA. Angry\textbackslash nB. Happy\textbackslash nC. Sad\textbackslash nD. Neutral",\\
    "What action causes the sharp sound?\textbackslash nA. Dropping a glass\textbackslash nB. Closing a door\textbackslash nC. Clapping hands\textbackslash nD. Stomping feet",\\
    "What sound marks the end of the clip?\textbackslash nA. A bell ringing\textbackslash nB. A fade out\textbackslash nC. A loud crash\textbackslash nD. An electronic beep"\\
  ]\\
\}\\

**Structure for Mismatch/Conflict:**\\
\{\\
  "is\_success": false,\\
  "questions": []\\
\}

\end{tcolorbox}

\begin{tcolorbox}[
    enhanced,
    colback=white,
    colbacklower=gray!20,
    colframe=black,
    boxrule=2pt,
    sharp corners,
    width=\textwidth,
    title={Answer Verification},
    colbacktitle=black,
    coltitle=white,
    fonttitle=\bfseries\sffamily\large,
    halign title=center,
    breakable,
    listing only,
    listing options={
        basicstyle=\ttfamily\footnotesize,
        columns=fullflexible,
        breaklines=false,
        literate={\textbackslash n}{{\textbackslash n}}2
    }
]

\# Role\\
You will assess the model's logic, factual grounding, and consistency based on specific domain rules.\\

\# Input Data\\
- **Domain Context:** \{L0\} > \{L1\} > \{L2\}\\
- **Caption (Ground Truth):** \{caption\}\\
- **Query:** \{query\}\\

\#\# Model A (Candidate 1)\\
- CoT: {model\_a\_cot}
- Answer: {model\_a\_response}\\

\#\# Model B (Candidate 2)\\
- CoT: {model\_b\_cot}\\
- Answer: {model\_b\_response}\\

\# General Instructions\\
1. **Consensus Check:** Do the models agree on the answer?\\
2. **Internal Logic Check:** Is the CoT self-consistent?\\
3. **Domain-Specific Adjudication (CRITICAL):**\\
   Apply the following rules strictly to determine the quality and winner:\\

   > START OF SPECIFIC RULES <\\
   \{INJECTION\_RULES\}\\
   > END OF SPECIFIC RULES <\\

\# Output Format (JSON)\\
Please output strictly in valid JSON format without markdown code blocks.\\
\{\{\\
    "consensus": true,\\
    "consistency\_with\_caption": "STRONG\_MATCH",\\
    "winner": "MODEL\_B",\\
    "reasoning": "string",\\
    "final\_action": "KEEP\_USING\_WINNER"\\
\}\}\\

**Valid values for fields:**\\
- "consensus": true | false\\
- "consistency\_with\_caption": "STRONG\_MATCH" | "WEAK\_MATCH" | "CONFLICT"\\
- "winner": "MODEL\_A" | "MODEL\_B" | "TIE" | "NONE"\\
- "final\_action": "KEEP\_USING\_WINNER" | "DISCARD"\\

\end{tcolorbox}

\subsection{Prompts for Dense Audio Captioning Evaluation}
\label{app:prompts_for_dense_caption}
In this section, we provide the prompts used in the dense audio captioning evaluation. Specifically, we include two types of prompts: (1) a caption generation prompt that instructs the model to produce comprehensive and semantically rich dense audio descriptions, and (2) a question answering prompt that takes the generated caption together with a downstream question as input to produce the final answer. The complete prompt formulations are presented to ensure transparency and reproducibility of the evaluation procedure.

\begin{tcolorbox}[
    enhanced,
    colback=white,
    colbacklower=gray!20,
    colframe=black,
    boxrule=2pt,
    sharp corners,
    width=\textwidth,
    title={Dense Caption Generation},
    colbacktitle=black,
    coltitle=white,
    fonttitle=\bfseries\sffamily\large,
    halign title=center,
    breakable,
    listing only,
    listing options={
        basicstyle=\ttfamily\footnotesize,
        columns=fullflexible,
        breaklines=false
    }
]

Please provide a comprehensive description of the input audio, covering all details.

\end{tcolorbox}

\begin{tcolorbox}[
    enhanced,
    colback=white,
    colbacklower=gray!20,
    colframe=black,
    boxrule=2pt,
    sharp corners,
    width=\textwidth,
    title={Question Answering},
    colbacktitle=black,
    coltitle=white,
    fonttitle=\bfseries\sffamily\large,
    halign title=center,
    breakable,
    listing only,
    listing options={
        basicstyle=\ttfamily\footnotesize,
        columns=fullflexible,
        breaklines=false
    }
]

\#\#\# System Role\\
You are an intelligent **Audio-Based Question Answering Assistant**. \\
You will receive a textual **[Audio Caption]** generated by a specialized audio analysis model that provides comprehensive details about the audio content.\\

\#\#\# Core Instructions\\
1. **Source Reliability**: The [Audio Caption] is your **sole source of truth**. It provides a detailed analysis of the audio, including speech content, speaker characteristics, background sounds, and emotional context.\\
2. **Mapping Task**: Your goal is to answer the [User Question] by extracting relevant evidence from the caption.\\
3. **Handling Limitations**: If the question asks for specific details that are not present in the caption, explicitly state that the information is not available in the audio analysis.\\

\#\#\# Thinking Process\\
Before answering:\\
1. **Analyze**: Carefully read the caption to understand all audio elements described.\\
2. **Verify**: Check if the caption contains sufficient evidence to support your answer.\\
3. **Reason**: Connect the observed audio elements to answer the question logically.\\

\#\#\# Input Data
\begin{verbatim}
[Audio Caption]: "{caption}"
[User Question]: "{question}"
\end{verbatim}

\#\#\# Output Format\\
Provide a direct and concise answer based on the audio caption. If the answer involves multiple-choice options, select the most appropriate one based on the evidence.\\

\end{tcolorbox}
\begin{figure}[h!]
  \centering
  \includegraphics[width=\textwidth]{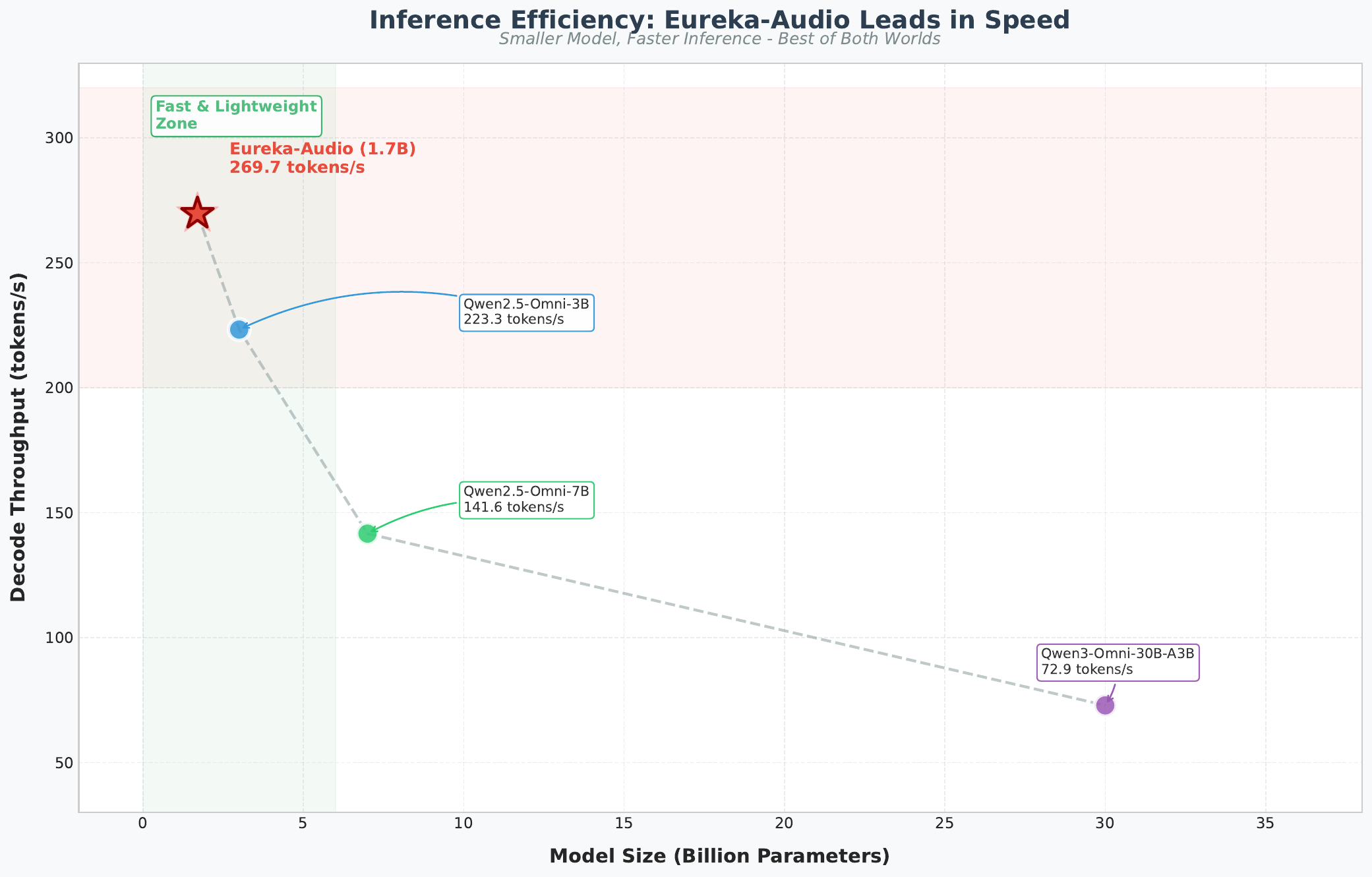}
  \caption{Decode throughput versus model size. Eureka-Audio-Instruct (1.7B) achieves the fastest inference at 269.7 tokens/sec, 3.7$\times$
  faster than Qwen3-Omni-A3B while being 17$\times$ smaller, highlighting its lightweight and efficient design.}
  \label{fig:speed_comp}
\end{figure}
\subsection{Decoding Throughput Comparison}
\label{sec:appendix_speed}
\paragraph{Hardware Setup.}
All experiments are conducted on a server equipped with dual
Intel Xeon Platinum 8468V processors
and NVIDIA H100 GPUs interconnected via NVSwitch.
Throughput results are measured using 8x H100 GPU.

\paragraph{Evaluation Protocol.}
We evaluate decoding throughput on 200 audio samples with a
maximum generation length of 2,000 tokens.
Throughput is computed as the average number of decoded tokens per second.
Figure~\ref{fig:speed_comp} compares the decode throughput across different model sizes. Eureka-Audio-Instruct (1.7B)
  achieves the highest throughput of 269.7 tokens/sec, which is 1.2$\times$ faster than Qwen2.5-Omni-3B, 1.9$\times$ faster
  than Qwen2.5-Omni-7B, and 3.7$\times$ faster than Qwen3-Omni-30B-A3B. With only 1.7B parameters, 17$\times$ smaller than
  the largest baseline—our model delivers superior inference speed, making it well-suited for on-device deployment on mobile
   phones and edge devices.

Eureka-Audio achieves the highest decoding throughput among all compared models,
demonstrating strong inference efficiency despite its compact parameter scale.

\newpage
\subsection{Qualitative Comparison of Dense Audio Captioning}
\label{app:qualitative_comparison}

In this section, we present a qualitative comparison of dense audio captions generated by three different models for an audio clip featuring a comedic dialogue from the sitcom \textit{Friends}.

\begin{tcolorbox}[
    enhanced,
    colback=white,
    colbacklower=gray!20,
    colframe=black,
    boxrule=2pt,
    sharp corners,
    width=\textwidth,
    title={Analysis of Eureka-Audio-Instruct (Ours)},
    colbacktitle=black,
    coltitle=white,
    fonttitle=\bfseries\sffamily\large,
    halign title=center,
    breakable
]
\small
The audio clip presents a fast-paced, comedic dialogue sequence, likely from a sitcom, characterized by a dry, studio-recorded acoustic environment with clear separation between the speakers. The scene opens with a male voice (Ross) delivering a line with a tone of exaggerated resignation and mild frustration: ``I want to quit the gym.'' A second male voice (Ross's partner, Chandler) responds with a flat, deadpan ``Okay.'' Ross continues, his voice sounding slightly more animated as he explains, ``Dave in the membership office handles quitters,'' followed by a brief pause and a sharp ``Excuse me,'' before asking, ``Are you a member?'' The second male responds with a curt, matter-of-fact ``No,'' and the first voice immediately delivers the punchline with a deadpan inflection: ``Sorry, members only.''

\medskip
The dynamic shifts abruptly as Ross repeats his initial complaint, ``I want to quit the gym,'' which is met with a stern, authoritative command from the second male: ``Be strong.'' This triggers a distinct, guttural vocalization from the second male---a deep, resonant groan or grunt of exertion (``Mmm!''). The audio then transitions to a new interaction where Ross asks, ``Are you a member of any gym?'' The second male responds with a dismissive, arrogant tone, ``No,'' and adds, ``And I'm not going to be, so you can save your little speech.'' The soundscape then introduces a third element: a female voice (Marisa) speaking in a polite, professional register, asking, ``Could you come in for a second?'' The second male hesitates with a low, ``Hmm,'' before pivoting to a new conversation with the female, asking, ``Hi, I'm Maria.'' Ross interjects, his voice sounding slightly more casual and conversational, asking, ``Hey, so did you go hit?'' The second male responds with a defensive, slightly sheepish tone, ``No. I almost did. Couldn't leave Ross there without a spotter.'' Ross then asks, ``So you joined the gym?'' The clip concludes with the second male stating, ``They're gonna take fifty bucks,'' delivered with a matter-of-fact, slightly cynical inflection. Throughout the sequence, the dialogue is crisp and foregrounded, with no significant background noise or music interfering with the speech.
\end{tcolorbox}

\begin{tcolorbox}[
    enhanced,
    colback=white,
    colbacklower=gray!20,
    colframe=black,
    boxrule=2pt,
    sharp corners,
    width=\textwidth,
    title={Analysis of Qwen3-Omni-30B-A3B-Instruct},
    colbacktitle=black,
    coltitle=white,
    fonttitle=\bfseries\sffamily\large,
    halign title=center,
    breakable
]
\small
It sounds like you're referencing a scene from the TV show \textit{Friends}! The clip you described is from the episode titled ``The One with the Blackout'' (Season 8, Episode 1). In this scene, Chandler tries to quit the gym, but he gets sidetracked and ends up joining it instead.

\medskip
Here's a breakdown of what happens:
\begin{enumerate}
    \item Chandler tells Joey that he wants to quit the gym.
    \item Joey tells him to talk to Dave, the guy in the membership office who handles quit requests.
    \item Chandler goes to talk to Dave, but it turns out Dave is only there to help people join the gym --- not quit.
    \item Chandler ends up getting persuaded to sign up instead, partly because of the gym's ``no commitment'' policy and the friendly, flirty interaction with a gym employee named Maria.
    \item Later, Chandler reveals that he didn't quit --- he actually joined. When Monica asks why he didn't just leave, he says, ``I couldn't leave Ross there without a spotter.''
\end{enumerate}

\medskip
It's a classic \textit{Friends} moment combining humor, character dynamics, and a little bit of gym culture satire. Chandler's reluctance to quit and his tendency to get distracted or talk himself into things is very true to his character. And yes --- the gym is ``no commitment'', but Chandler still ends up stuck with a membership! Classic Chandler.
\end{tcolorbox}

\begin{tcolorbox}[
    enhanced,
    colback=white,
    colbacklower=gray!20,
    colframe=black,
    boxrule=2pt,
    sharp corners,
    width=\textwidth,
    title={Analysis of Qwen3-Omni-30B-A3B-Captioner},
    colbacktitle=black,
    coltitle=white,
    fonttitle=\bfseries\sffamily\large,
    halign title=center,
    breakable
]
\small
The audio clip opens inside a bustling gymnasium, where the constant, rhythmic clatter of weights and equipment underscores a comedic exchange. A male voice, exasperated, announces, ``I want to quit the gym,'' prompting a calm, matter-of-fact response from a second male: ``Okay. Dave in the membership office handles quitters.'' Immediately, the first man calls out, ``Excuse me! Are you a member?'' His question is met with a blunt ``No,'' followed by the sharp retort, ``Sorry. Members only.'' The initial speaker repeats his intent with increased urgency, ``I want to quit the gym!''---only to be met with the gym's slogan, ``Be strong!'' This is immediately punctuated by a cartoonish ``Woo!'' and a burst of audience laughter, signaling a comedic beat.

\medskip
The scene shifts, and the same male voice, now slightly exasperated, asks, ``Are you a member of any gym?'' Another ``No'' follows, with the reply, ``And I'm not going to be, so you can save your little speech.'' A female voice then interjects, ``Could you come here for a second?'' Her polite introduction, ``Hi, I'm Maria,'' is delivered with a friendly, inviting tone, but is met with a hesitant, awkward ``Hi.'' This exchange is again followed by audience laughter, underscoring the comedic timing.

\medskip
Throughout, the gym's ambient sounds---weights clanking, equipment humming, and a constant crowd murmur---provide an authentic backdrop. The recording is clean and modern, with each voice clearly separated and audience reactions well-captured. The dialogue and comedic timing are typical of a sitcom sketch. The characters, voices, and setting are consistent with the style and cast of the American TV show \textit{Friends}, with the final segment featuring Ross and Phoebe in a gym conversation that humorously subverts expectations.

\medskip
\textbf{Summary:} In summary, this audio clip presents a lively, comedic gym scene from a modern American sitcom, characterized by clear dialogue, energetic gym ambiance, and audience laughter. The skit plays on gym culture tropes, featuring a failed attempt to quit, playful banter, and a punchline involving membership fees.
\end{tcolorbox}

\end{document}